\documentclass[aps,prb,twocolumn,showpacs,floatfix]{revtex4}
\usepackage{graphicx}
\usepackage{dcolumn}
\usepackage{amsmath,amssymb,mathrsfs}
\usepackage{hhline}
\usepackage{wasysym}
\newcommand{\be}{\begin{equation}}
\newcommand{\ee}{\end{equation}}
\newcommand{\bea}{\begin{eqnarray}}
\newcommand{\eea}{\end{eqnarray}}
\newcommand{\re}{{\rm Re\;}}
\newcommand{\im}{{\rm Im\;}}
\newcommand{\G}{{\cal G}}
\newcommand{\ds}{\displaystyle}
\newcommand{\1}{{1\hspace*{-0.5ex} \textrm{l} \hspace*{0.5ex}}}
\renewcommand{\ds}{\displaystyle}
\newcommand{\nn}{\nonumber \\}
\newcommand{\sgnx}{{\rm sgn}(x)\;}
\newcommand{\B}{{\cal B}}
\newcommand{\K}{{\cal K}}

\begin{document}
\title{Dynamical correlations in the spin-half two-channel Kondo model}
\author{A.\ I.\ T\'oth and 
 G.\ Zar\'and
}
\affiliation{
Theoretical Physics Department, Institute of Physics, 
Budapest University of Technology and Economics, H-1521 Budapest, Hungary
}

\date{\today}
\begin{abstract}
Dynamical correlations of various local operators are studied in the 
spin-half two-channel Kondo (2CK) model in the presence of channel anisotropy or external magnetic
field. A conformal field theory-based scaling approach is used to predict the
analytic properties of various spectral functions in the vicinity of the two-channel Kondo
fixed point. These analytical results compare well with highly accurate
density matrix  numerical renormalization group results. The universal cross-over functions
interpolating between channel-anisotropy or magnetic field-induced Fermi
liquid regimes  and the two-channel Kondo, non-Fermi liquid regimes are determined numerically.
The boundaries of the real 2CK scaling regime are found to be rather
restricted, and to depend both on the type of the perturbation and on the
specific operator whose  correlation function is studied.  In a small magnetic field, a
universal resonance is observed in the local fermion's spectral function. The
dominant superconducting instability appears in the composite superconducting channel.
\end{abstract}
\maketitle

\section{Introduction}

Deviations from Fermi liquid-like behavior observed e.g.\ in the
metallic state of high-temperature cuprate 
superconductors,\cite{Lee_07,Carlson_04} or in heavy fermion
systems\cite{Lohneysen_07,Coleman_06} 
prompted 
physicists  
to look for new non-Fermi liquid (NFL) compounds.
So far a large number of such exotic compounds has been found
and investigated. 
In these systems electrons remain incoherent 
down to very low temperatures and the usual Fermi liquid description 
breaks down. To our current understanding, NFL 
physics may arise in many different ways: 
it can occur due to some local dynamical quantum fluctuations 
often described by quantum impurity models,\cite{Cox_97,Si_04,Vojta_06} 
it can also be attributed to the presence of the quantum fluctuations of an
order parameter or some collective modes, as is the case in the vicinity of 
many quantum phase transitions,\cite{Si_04,QPT_review_Vojta_03} 
or 
for  
the prototypical example of a Luttinger liquid,\cite{Luttinger,Bockrath_99,Ishii_03,Singer_05,Dora_07} where
electrons are totally disintegrated into collective excitations of the
electron gas.
%
%
NFL physics 
can also appear as a consequence of disorder like e.g.\ in disordered 
Kondo alloys.\cite{Milovanovic,Dobrosavljevic} 

In this paper we study a variant of the overscreened multi-channel Kondo model: 
the spin-$\frac{1}{2}$, two-channel Kondo (2CK) model, which is the simplest
prototypical example of non-Fermi liquid quantum impurity models. 
This model has first been introduced by Nozi\`eres and Blandin,\cite{Nozieres_81} 
and since has been proposed to
describe a variety of systems including
dilute heavy fermion compounds,\cite{Cox_97}
tunneling impurities in disordered metals and doped 
semiconductors.\cite{Katayama_87,Jan_97,Steglich_05} 
More recently, the 2CK state has been observed in a very controlled way in a double dot system
originally proposed by 
Oreg and 
Goldhaber-Gordon.\cite{Oreg_03,Potok_07}\footnote{For further theoretical
  studies see Ref.-s \onlinecite{Pustilnik_04,Schiller_04,2ck_cond}.}

The two-channel Kondo model consists of a spin-$\frac{1}{2}$, local moment
which is coupled through antiferromagnetic exchange interactions 
to two channels of conduction electrons. Electrons in both  
channels try to screen the impurity spin.
If the coupling of the spin to 
one of the
channels is 
stronger than to the other
then electrons in the more strongly coupled channel
screen the spin, while the other channel becomes
decoupled. 
However, 
for equal exchange couplings, the competition between the two
channels leads to overscreening and results in a non-Fermi liquid behavior:
Among others, 
it is characterized by a non-trivial 
zero temperature residual entropy, 
a square root-like temperature dependence of the differential conductance, 
a logarithmic divergence of the spin susceptibility
and the linear specific heat coefficient at low temperatures.\cite{Cox_97}  
This unusual and fragile ground state cannot be described within the
framework of 
Nozi\`eres' 
Fermi liquid theory.\cite{Nozieres_74} 

Being a prototypical example of non-Fermi liquid models, the 
two-channel Kondo model (2CKM) has already been investigated with 
a number of methods. 
These 
include non-perturbative techniques
like the Bethe Ansatz, which gives full account of the 
thermodynamic properties,\cite{Andrei_84,Tsvelick_85} 
boundary conformal field theory,\cite{Affleck_Ludwig_91} which describes the vicinity of
the fixed points, and numerical renormalization group (NRG)
methods.\cite{NRG_ref}
Furthermore other less powerful
approximate methods such as the Yuval-Anderson approach,\cite{Zimanyi_88} Abelian 
bosonization,\cite{Emery_Kivelson} large-$f$ expansion,\cite{Gan_93,Zarand_96} and 
non-crossing approximation\cite{Cox_93} have also been used to study the 2CKM successfully. 

Rather surprisingly, despite this extensive work, 
very little is known about {\em dynamical} correlation functions 
such as the spin susceptibility, local charge and superconducting
susceptibilities. Even the detailed properties of the $T$-matrix,
essential to understand elastic and inelastic scattering in this non-Fermi
liquid case,\cite{Laci_inel} have 
only been computed earlier using conformal
field theory (which is rather limited in energy range) 
and by the non-crossing approximation (which is not well-controlled 
and is unable to describe the Fermi liquid cross-over).\cite{Affleck_93,Kroha_97} 
It was also possible to compute some of the dynamical correlation 
functions in case of  extreme spin anisotropy using 
Abelian bosonization results,\cite{Emery_Kivelson} though these 
calculations reproduce only partly the generic features 
of the spin-isotropic model.\cite{Sengupta_Georges} 
Local correlations in the Anderson model 
around the non-Fermi liquid fixed point have already been 
investigated with the use of NRG, although in the absence of channel
anisotropy and magnetic field.\cite{Suzuki_97} 
However, a thorough and careful NRG analysis of the 
$T=0$ temperature $T$-matrix of the 2CKM has been carried out only 
very recently,\cite{2ck_cond,Laci_inel}  
and the $T\ne0$ analysis still needs to be done.

The main purpose of this paper is to fill this gap 
by giving a comprehensive 
analysis of the local correlation functions at zero temperature using the
numerical renormalization group approach. 
However, in the vicinity of the rather
delicate two-channel Kondo fixed point, the conventional NRG method fails 
and its further developed version, the density matrix NRG (DM-NRG)\cite{DMNRG} 
needs to be applied. Furthermore, a rather
large number of multiplets must be kept to achieve good accuracy.
We have therefore implemented a modified version of the recently developed spectral sum
conserving DM-NRG method, where we use non-Abelian symmetries in a
flexible way to compute  the real and the imaginary parts of various local
correlation functions.\cite{DMNRG_unpublished} 

To identify the relevant perturbations around the NFL fixed point we apply the machinery
of boundary conformal field theory. 
Then we systematically study how the vicinity of fixed points and the introduction of relevant 
perturbations such as a finite channel anisotropy or a finite magnetic field 
influence the form of the dynamical response functions at zero temperature. 
We mainly focus on the strong coupling regime of the 2CK model and 
the universal cross-over functions in the proximity of this 
region induced by an external magnetic field or channel anisotropy. 
We remark that these cross-over functions, describing the cross-over
from the non-Fermi liquid fixed point to a Fermi liquid fixed point,
as well as the response functions can currently be computed reliably 
at all energy scales only  with NRG. 
However, we shall be able to use the results of boundary conformal
field theory, more precisely, the knowledge of the operator content of the
two-channel Kondo fixed point and  the scaling dimensions of the various
perturbations around it, to make very general statements 
on the analytic properties of the 
various  cross-over and spectral functions.

We shall devote special attention to superconducting fluctuations. It has 
been proposed that unusual superconducting states observed in some incoherent 
heavy fermion compounds could also emerge as a result of local superconducting 
correlations associated with 
two-channel Kondo
physics.\cite{Jarrell_96,Cox_97,Coleman_98} Here we investigate 
some possible superconducting order parameters consistent with the conformal
field 
theoretical predictions, and find that the dominant instability emerges in
the so-called {\em composite superconducting channel}, as it was proposed 
by Coleman {\em et al}.\cite{Coleman_98}

The paper is organized as follows. In Section \ref{sec:ham_sym} starting 
from the one-dimensional, continuum formulation of the 2CKM we connect it
to a dimensionless approximation of it suited to our DM-NRG calculations.
We also provide the symmetry generators used in the conformal field theoretical and
DM-NRG calculations. In Section \ref{sec:NFL_fixed_point} we use 
boundary conformal field theory to classify the boundary highest-weight fields of the 
electron-hole symmetrical 2CKM by
their quantum numbers and identify the relevant perturbations around the 2CK
fixed point. Based on this classification the fields are
then expanded in leading order in terms of the operators of the free theory. 
In Section \ref{sec:nrg} we describe the technical details of our DM-NRG calculations.
In Sections \ref{sec:fermion}, \ref{sec:spin} and \ref{sec:superC} we study the real and the
imaginary parts of the retarded Green's functions of the local fermions, the
impurity spin and the local superconducting order parameters. In
each of these sections we first discuss the analytic forms of the
susceptibilities in the asymptotic regions of the two-channel and single channel Kondo
scaling regimes, as they follow from scaling arguments. Then
we confirm our predictions by demonstrating how the expected corrections due
to the relevant perturbations and the leading irrelevant operator present 
themselves in the DM-NRG data. Furthermore we determine the boundaries of the 2CK scaling regimes
and derive universal scaling curves connecting the FL and NFL fixed points 
for each operator under study.
Finally, our conclusions are drawn in Section \ref{sec:fin}.

\section{Hamiltonian and symmetries\label{sec:ham_sym}}

The two-channel Kondo model consists of 
an impurity with a magnetic moment $S=\frac{1}{2}$ 
embedded into a Fermi liquid (FL) of two types of electrons (labeled by 
the flavor or channel indices $\alpha=1,2$), and interacting with them 
through a simple exchange interaction,  
\bea
\mathscr{H}  &=&  \sum_{\alpha,\mu} \int_{-{D_F}}^{{D_F}} {\rm d}k  \; k\;  
c^\dagger_{\alpha,\mu}(k)c^{}_{\alpha,\mu}(k)\;
\label{eq:cont_Kondo}
\\
&+& \sum_\alpha \sum_{\mu,\nu} \frac{J_{\alpha}}2    \int_{-{D_F}}^{{D_F}} {\rm d}k \int_{-{D_F}}^{{D_F}}
{\rm d}k' \; 
 {\vec S} \;c^\dagger_{\alpha\mu}(k)  {\vec\sigma}_{\mu\nu} c^{}_{\alpha\nu}(k') 
\;.
\nonumber
\eea
Here $c^\dagger_{\alpha,\mu}(k)$ creates an electron of flavor $\alpha$ 
in the $l=0$ angular momentum channel  with spin $\mu$ and radial 
momentum $k$ measured from the Fermi momentum. 
In the Hamiltonian above we allowed for a channel anisotropy of the
couplings, $J_1 \ne 
J_2$, and denoted the Pauli matrices by  $\vec\sigma$. 
In the first, kinetic term, we assumed a spherical Fermi surface and 
linearized the spectrum of the conduction electrons, $\xi(k)\approx v_F k =
k$, but these assumptions are not crucial: 
Apart from irrelevant terms in the Hamiltonian, 
our considerations below carry over to essentially any 
local density of states with electron-hole symmetry.
The fields $c^\dagger_{\alpha,\mu}(k)$ are normalized to satisfy the anticommutation relations
\bea
\left\{c^\dagger_{\alpha,\mu}(k),c^{}_{\beta,\nu}(k^\prime)\right\}=
\delta_{\alpha,\beta}
\delta_{\mu,\nu} \delta\left(k-k^\prime\right)
 \;,
\eea
and therefore the couplings $J_\alpha$ are just the dimensionless couplings,
usually defined in the literature. Since we are
interested in the low-energy properties of the system, an energy cut-off ${D_F}$ is
introduced for the kinetic and the interaction energies. 
In heavy fermion systems, this large energy
scale is in the range of the Fermi energy, ${D_F}\sim E_F$, while for quantum
dots, it is of the order of the single particle level spacing
 of the dot, $\delta\epsilon$ or its charging energy, $E_C$, whichever is smaller.

\begin{table*}[htb]
  \begin{tabular}{cccc}
    \hhline{====}
    Symmetry group&&Generators&\\
    \hline
    $\textrm{SU}_{C\alpha}(2)$&${\rm C}^{+}_{\alpha}= \ds{\sum_{n=0}^{\infty}}(-1)^nf^{\dagger}_{n,\alpha,\uparrow}f^{\dagger}_{n,\alpha,\downarrow}\;,$&
    ${\rm C}^z_\alpha=\frac{1}{2}\ds{\sum_{n=0}^{\infty}}~\ds{\sum_{\mu}}
    \left(f^{\dagger}_{n,\alpha,\mu}f_{n,\alpha,\mu}^{}-1\right)\;,$&${\rm
      C}^{-}_\alpha= {\rm C}^{+\dagger}_\alpha$\\
    $\textrm{SU}_S(2)$&&$
    {\vec {\rm J}}={\vec S}+\frac{1}{2}\ds{\sum_{n=0}^{\infty}}
    \ds{\sum_{\alpha,\mu,\nu}}
    f^{\dagger}_{n,\alpha,\mu}{\vec\sigma}^{}_{\mu\nu}f^{}_{n,\alpha,\nu}$&\\
    \hhline{====}
    \end{tabular}
  \caption{Generators of the used symmetries for the two-channel Kondo model
    computations. Sites along the Wilson chain  are
    labeled by $n$ whereas $\alpha$ and $\mu,\nu$ are the channel and spin indices,
    respectively.}
  \label{tab:symmetry_generators}
\end{table*}

The Hamiltonian above possesses various symmetries. To see it, it is worth
 to introduce
the left-moving fermion fields, 
\be
\psi_{\alpha,\mu}(x) \equiv \int^{D_F}_{-{D_F}} dk\; e^{-i k x} c^{}_{\alpha,\mu}(k)\;,
\ee
and 
 to rewrite 
the Hamiltonian as 
\begin{multline}
\mathscr{H} =\sum_{\alpha,\mu} \int \frac {{\rm d}x}{2\pi }  \;  
\psi^\dagger_{\alpha,\mu}(x)\; i \partial_x \psi^{}_{\alpha,\mu}(x)
\\
+\sum_\alpha \frac {J_\alpha} 2 \vec S \psi^\dagger(0)\vec \sigma\psi(0)\;.
\label{eq:H_psi}
\end{multline}  
Then the total spin operators $\mathscr{J}^{i}$ defined as 
\bea
\mathscr{J}^{i} &\equiv& S^i + \int \frac {{\rm d}x}{2\pi}\; {J}^{i}(x)\;,
\\
{J}^{i}(x)
&\equiv&\frac 12 \ds{\sum_{\alpha}} 
:\psi^\dagger_{\alpha}(x) {\sigma}^i\psi^{}_{\alpha}(x):\;
\eea
commute with the Hamiltonian and satisfy the standard
SU(2) algebra,
\be
\left[ \mathscr{J}^i,\mathscr{J}^j \right]  = i  \epsilon^{ijk}\mathscr{J}^k\;.
\ee
 In the previous equations we suppressed spin indices and introduced
the normal ordering $:...:$ with respect to the non-interacting Fermi sea.
In a similar way we can define the ``charge spin'' density 
operators, for the channels  $\alpha=1,2$ as
\bea
{C}_\alpha^{z}(x)
&\equiv&\frac 12 
:\psi^\dagger_{\alpha}(x)\psi^{}_{\alpha}(x):\;
\nonumber
\\
{C}_\alpha^{-}(x) 
&\equiv&  
\psi^{}_{\alpha\uparrow}(x) \psi^{}_{\alpha\downarrow}(x), \phantom{nn}
{C}_\alpha^{+}(x) \equiv 
\psi^{\dagger}_{\alpha\downarrow}(x)
\psi^{\dagger}_{\alpha\uparrow}(x)  
\;,
\nonumber
\\
{C}_\alpha^{\pm}(x)
&\equiv& {C}_\alpha^{x}(x) \pm i\; {C}_\alpha^{y}(x)
\;,
\eea
and the corresponding symmetry generators
\be
\mathscr{C}_\alpha^{i} \equiv  \int \frac {{\rm d}x}{2\pi}\; {C}_\alpha^{i}(x)\;
\phantom{nnn}(i=x,y,z)\;.
\ee
The generators $\mathscr{C}_\alpha^i$, 
which are related to the
electron-hole symmetry,\cite{Jones_87}  
satisfy the same SU(2) algebra as the  $\mathscr{J}^i$-s, 
\begin{eqnarray} 
\left[\mathscr{C}_\alpha^i,\mathscr{C}_\beta^j\right] & =& i \delta_{\alpha\beta} \;
\epsilon^{ijk} \mathscr{C}_\beta^k\;,
\end{eqnarray}
and they also commute with the Hamiltonian, Eq.~\eqref{eq:H_psi}.  
Thus the Hamiltonian $\mathscr{H}$
has a symmetry $\textrm{SU}_{C1}(2)\times
\textrm{SU}_{C2}(2)\times\textrm{SU}_S(2)$ in the charge and spin sectors
for arbitrary couplings, $J_{\alpha}$. 

To perform NRG calculations, we use the following approximation
of the dimensionless Hamiltonian,\cite{NRG_ref}
\begin{multline}
\frac{2\mathscr{H}}{{D_F}\left(1+\Lambda^{-1}\right)}\approx
\ds{\sum_{\alpha}} \ds{\sum_{\mu,\nu}} 
\frac {{\tilde J}_{\alpha}}2\;
{\vec S}
f_{0,\alpha,\mu}^{\dagger}{{\vec\sigma}_{\mu\nu}}f_{0,\alpha,\mu}^{}\\
+\ds{\sum_{n=0}^\infty}\sum_{\alpha,\mu,\nu} t_n 
(f_{n,\alpha,\mu}^{\dagger} f_{n+1,\alpha,\mu}^{} + h.c.)\;,
\label{eq:Kondo_nrg}
\end{multline}
with $\Lambda$ a discretization parameter
and  ${\tilde J}_{\alpha}=
4J_{\alpha}/(1+\Lambda^{-1})$. The operator, 
$f_0$ creates an electron right at the impurity site, and can be expressed 
as 
\bea
f_{0,\alpha,\mu}=\frac{1}{\sqrt{2{D_F}}}\ds{\int^{{D_F}}_{-{D_F}}}{dk}\;
c_{\alpha,\mu}(k)
\;.
\label{eq:f0_acom}
\eea
The Hamiltonian  
Eq.\ (\ref{eq:Kondo_nrg}) is also called the Wilson chain: it 
describes electrons hopping along a semi-infinite chain with a 
hopping amplitude $t_n\sim \Lambda^{-n/2}$, and interacting with 
the impurity only at site 0. In the NRG procedure, this Hamiltonian is
diagonalized iteratively, and its spectrum is used to compute the
spectral functions of the various operators.\cite{NRG_ref}

We remark that the Wilson Hamiltonian is not identical 
to $\mathscr{H}$, since some terms are neglected along its 
derivation.\cite{NRG_ref} Nevertheless, similar 
to $\mathscr{H}$, 
the Wilson Hamiltonian also possesses the symmetry $\textrm{SU}_{C1}(2)\times
\textrm{SU}_{C2}(2)\times\textrm{SU}_S(2)$ for arbitrary
$J_{1}$ and $J_2$ couplings.\cite{Jones_87} 
The corresponding symmetry generators have been enumerated in   
Table~\ref{tab:symmetry_generators}. We can then use these
symmetries to label every multiplet in the Hilbert space 
and every operator multiplet by the eigenvalues
$\vec{\mathscr{J}}^{\;2}=j(j+1)$ 
and 
$\vec{\mathscr{C}}_\alpha^{\;2}=c_\alpha(c_\alpha +1)$. 
Throughout this paper, we shall
use these quantum numbers to 
classify states and operators.

In the presence of a magnetic field, i.e., when a term
\footnote{Throughout the paper we use units of $\hbar=k_B=v_F=1$.} 
\bea
H_{magn}=- g\mu_B B\; S^z
\eea
is added to $\mathscr{H}$,
the symmetry of the system breaks down to
$\textrm{SU}_{C1}(2)\times\textrm{SU}_{C2}(2)\times\textrm{U}_S(1)$,  
with the symmetry $\textrm{U}_S(1)$ corresponding to the conservation of the 
$z$-component of the spin, $\mathscr{J}^z$ (see Table~\ref{tab:symmetry_generators}).
In the rest of the papers we shall use units where we set $g\mu_B\equiv 1$.

\section{The non-Fermi liquid fixed point and its operator content}
\label{sec:NFL_fixed_point}

\begin{figure}[ht]
  \includegraphics[width=0.9\columnwidth,clip]{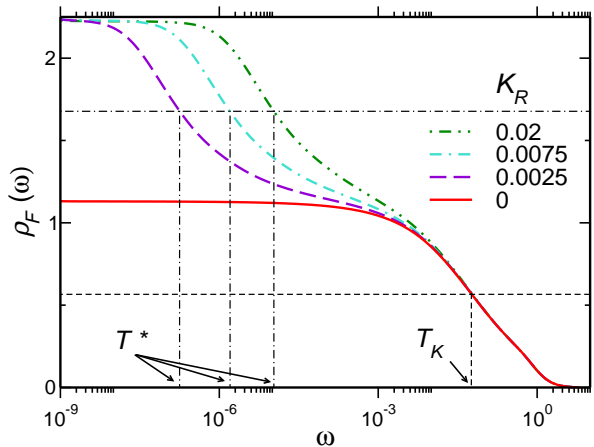}
  \caption{(color online) Spectral function $\varrho_F$ of the composite
    fermion operator,  $F_{0,1,\uparrow}$ 
    as a function of $\omega$, and the definition of the scales $T_K$ and $T^*$.
    $T_K$ is defined by the relation $\varrho_{F}(\omega=T_K,T=0,K_R=0)\equiv
    \frac 1 2 \varrho_{F}(\omega=0,T=0,K_R=0)$. 
%
%
For non-zero $K_R$
the scale $T^*$ is defined through 
   $\varrho_{F}(\omega=T^\ast,T=0,K_R)\equiv
    \frac{3}{4}\varrho_{F}(\omega=0,T=0,\left|K_R\right|)$. 
  }
  \label{fig:rho_F}
\end{figure}

For $J_1=J_2=J$ and in the absence of an external magnetic field, 
the Hamiltonian, $\mathscr{H}$ possesses a dynamically 
generated energy scale, the so-called Kondo temperature,
$$
T_K\approx {D_F} \; e^{-1/J}.
$$ 
The definition of $T_K$ is
somewhat arbitrary. In this paper, 
$T_K$ shall be defined as the energy $\omega$ at which for $J_1=J_2$ 
the 
spectral function of the composite fermion drops to half of its value 
assumed at $\omega=0$ (for further details see the end of this Section 
and Fig.\ \ref{fig:rho_F}). 
For $B=0$ and $J_1=J_2$, below this energy scale the physics is governed by
the so-called two-channel Kondo fixed point. 

The physics of the two-channel Kondo fixed point and its vicinity can be
captured using conformal field theory.
The two-channel Kondo finite size spectrum  and its
operator content has first been obtained using boundary conformal field theory by 
Affleck and Ludwig.\cite{Affleck_Ludwig_91} 
However, instead of  charge SU(2) symmetries, Affleck and Ludwig used 
flavor SU(2) and 
charge U(1) symmetries to obtain the fixed point
spectrum.\cite{Affleck_Ludwig_91} The use of charge SU(2) symmetries, however, has a clear advantage 
over the flavor symmetry when it comes to performing NRG calculations: While
the channel anisotropy violates the flavor symmetry, it does not violate the charge SU(2) symmetries.  
Therefore, even in the channel anisotropic case, we have three commuting SU(2) symmetries. 
If we switch on a local magnetic field, only the spin SU(2) symmetry is reduced to its U(1) subgroup. 
Using charge symmetries allows thus for much
more precise calculations, and in fact, using them is absolutely necessary to obtain  
satisfactorily accurate spectral functions, especially in the presence of a magnetic field. 

To understand the fixed point spectrum and the operator content of the 2CKM, 
let us outline the boundary conformal field theory in this 
$\textrm{SU}_{C1}(2)\times\textrm{SU}_{C2}(2)\times\textrm{SU}_S(2)$ 
language. First, we remark that the spin density operators,
$J^i(x)$ satisfy the SU(2)$_{k=2}$ Kac-Moody algebra of level $k=2$, 
\bea
\left [ J^i(x),J^j(x')\right]& =& \frac k 2 \delta^{ij} \;\delta^\prime(x-x')
\nonumber \\
&+& i \;2\pi\;\delta(x-x') \epsilon^{ijk}J^k(x)\;, 
 \eea
while the charge density operators, $C_\alpha^i(x)$ defined in the previous section 
satisfy the Kac-Moody algebra of level $k=1$: 
\bea
\left [ C_\alpha^i(x),C_\beta^j(x')\right] &=& \frac k 2 \delta^{ij}\delta_{\alpha\beta} \;\delta^\prime(x-x')
\nonumber \\
&+& i \;2\pi\;\delta_{\alpha\beta}\delta(x-x') \epsilon^{ijk}C_\alpha^k(x)\;. 
\nonumber 
\eea
We can use these current densities and the coset
construction to write the kinetic part of the  Hamiltonian as
\bea
\mathscr{H}_0&=& \mathscr{H}_{C1} +\mathscr{H}_{C2} + \mathscr{H}_{S} +
\mathscr{H}_{I}\;,
\label{eq:H_sug1}
\\
\mathscr{H}_{C\alpha} &=& \frac 1 3 \int \frac {{\rm d}x}{2\pi} : \vec C_\alpha (x)\vec C_\alpha (x):\;,
\nonumber 
\\
\mathscr{H}_{S} &=& \frac 1 4 \int \frac {{\rm d}x}{2\pi} : {\vec J} (x){\vec J} (x):\;.
\nonumber 
\eea
In $\mathscr{H}_0$, the first two terms describe the charge sectors, and have central charge 
$c=1$, while $\mathscr{H}_{S}$ describes the spin sector, and has central 
charge $c=3/2$. The last term corresponds to the coset space, and must have 
central charge $c=1/2$, since the free  fermion model has central
charge $c=4$, 
corresponding to the four combinations of spin and channel quantum numbers.
 This term can thus be identified as the Ising model, 
having primary fields  $\1,\sigma,\epsilon$ with 
scaling dimensions $0,1/16,1/2$,
respectively. We can then carry out the conformal embedding in the usual way, by comparing
the finite size spectrum of the free Hamiltonian with that of
Eq.~\eqref{eq:H_sug1}, and identifying the allowed primary fields in the product
space. The the fusion rules obtained this way are listed on the left side
of Table~\ref{tab:free_primaries}. The finite size spectrum 
at the two-channel Kondo fixed point can be derived by fusing with the impurity
spin (which couples to the spin sector only), following the operator product
expansion of the 
Wess--Zumino--Novikov--Witten model, 
$1/2\otimes 0\rightarrow1/2$, $1/2\otimes 1/2\rightarrow 0\oplus1$,
$1/2\otimes 1\rightarrow1/2$ (see RHS of Table~\ref{tab:free_primaries}). 
Finally, the operator content of the fixed point can be found by performing a
second fusion with the spin. The results
of this double fusion are presented in Table~\ref{tab:conf_weights}.
In Table~\ref{tab:conf_weights} the leading irrelevant operator, ${\vec {\cal J}}_{-1}{\vec\phi}_{s}$, 
is also included. Although it is not a primary field,\cite{Affleck_Ludwig_91} 
close to the 2CK fixed point, 
this operator will also
 have impact on the form 
the correlation functions.

\renewcommand{\arraystretch}{1.2}
\begin{table}[t]
  \begin{tabular}{cccc|c}
    \hhline{=====}
    $c_1$&$c_2$&$j$&$I$&$E_{\rm free}$\\
    \hline
    0&0&0&$\1$&0\\
    $\frac{1}{2}$&0&$\frac{1}{2}$&$\sigma$&$\frac{1}{2}$\\
    0&$\frac{1}{2}$&$\frac{1}{2}$&$\sigma$&$\frac{1}{2}$\\
    $\frac{1}{2}$&$\frac{1}{2}$&1&$\1$&1\\
    $\frac{1}{2}$&$\frac{1}{2}$&0&$\epsilon$&1\\
    \hhline{=====}
  \end{tabular}\hspace{1cm}
 \begin{tabular}{cccc|c}
    \hhline{=====}
    $c_1$&$c_2$&$j$&$I$&$E_{\rm 2CKM}$\\
    \hline
    0&0&$\frac{1}{2}$&$\1$&0\\
    $\frac{1}{2}$&0&0&$\sigma$&$\frac{1}{8}$\\
    0&$\frac{1}{2}$&0&$\sigma$&$\frac{1}{8}$\\
    $\frac{1}{2}$&$\frac{1}{2}$&$\frac{1}{2}$&$\1$&$\frac{1}{2}$\\
    $\frac{1}{2}$&$0$&$1$&$\sigma$&$\frac{5}{8}$\\
    0&$\frac{1}{2}$&$1$&$\sigma$&$\frac{5}{8}$\\
    $\frac{1}{2}$&$\frac{1}{2}$&$\frac{1}{2}$&$\epsilon$&1\\
    \hhline{=====}
  \end{tabular}
  \caption{Left: Primary fields and the corresponding finite size energies  
    at the free fermion fixed point for  anti-periodic boundary
    conditions. States are classified according to the group 
    $\textrm{SU}_{C1}(2)\times\textrm{SU}_{C2}(2)\times\textrm{SU}_S(2)$
    and the Ising model. The excitation energies $E_{\rm free}$ are given in  units of
    $2\pi/L$,  with $L$ the size of the chiral fermion system.
    Right: Finite size spectrum  at the two-channel Kondo fixed point.  
}
  \label{tab:free_primaries}
\end{table}

What remains is to identify the scaling operators in terms of the operators of the
non-interacting theory. In general, an operator of the non-interacting theory
can be written as an infinite series in terms of the scaling operators
and their descendants. Apart from the Ising sector, which is hard to 
identify, we can tell by looking at the various quantum numbers of an operator
acting on the Wilson chain, 
which primary fields could be present in it. In this way, we can
identify, e.g. $\vec \phi_s$ as the spin operator $\vec S$. 
Thus 
the spin operator can be expressed as 
\be
\vec S = A_s\; \vec \phi_s + \dots\;
\ee
where the dots stand for all the less relevant operators that are present in the
expansion of $\vec S$, 
and some high-frequency portions which are not properly captured in the 
expansion above. 
The weight, $A_s$, can be determined from matching the 
decay of the spin-spin correlation function at short and long times. This
way we end up with $A_s\sim 1/\sqrt{T_K}$.

We remark that there are infinitely many operators that contain the scaling 
fields in their expansion. As an example, consider the operators 
$\phi^{\tau\sigma}_{\psi1}$.
 Here the label $\sigma=\{\uparrow,\downarrow\}$ refers
to the spin components of a $j=1/2$ spinor, while $\tau=\pm$ refer to 
the charge spins of a charge $c=1/2$ spinor. To identify the corresponding
operator on the Wilson chain, 
we  first note that $f^\dagger_{0,1,\sigma}$
transforms as a spinor under spin rotations. It can easily be seen that 
the operator  $\tilde
f^\dagger_{0,1}\equiv i\sigma_y f_{0,1}$ also transforms as a
spinor. 
We can then form a four-spinor 
out of these operators,  
$\gamma_1\equiv\{ f^\dagger_{0,1,\sigma}, \tilde f^\dagger_{0,1,\sigma}\}$. It
is easy to show 
 that $\gamma_1$ transforms as a spinor 
under ${\rm SU}_{C1}(2)$ rotations as well, thus $\phi^{\tau\sigma}_{\psi1}$ could be
identified as $\gamma_1=\{ f^\dagger_{0,1,\sigma}, \tilde f^\dagger_{0,1,\sigma}\}$. 

However,  
we can construct another operator, 
$F^\dagger_{1}\equiv f^\dagger_{0,1} \vec S \vec \sigma $ and its
counterpart, $\tilde
F^\dagger_{1}\equiv i\sigma_y F_{1}$, and form a four-spinor out of them: 
$\Gamma_1\equiv\{ F^\dagger_{1,\sigma}, \tilde
F^\dagger_{1,\sigma}\}$. 
This operator has the same quantum numbers 
as $\gamma_1$, and in fact, both operators' expansion  contains
$\phi^{\tau\sigma}_{\psi1}$. 

\begin{table*}[t]
\label{tab:conf_weights}
  \begin{tabular}{cccccccc}
    \hhline{=======}
    $c_1$&$c_2$&$j$&$I$&$x^{2CK}$ &scaling& corresponding
    operators
    \\
    &&&&&operators & 
    \\
    \hline
    0&0&1&$1$&$\frac{1}{2}$&${\vec \phi}_s$& $\vec S$
    \\
    $\frac{1}{2}$&0&$\frac{1}{2}$&$\sigma$&$\frac{1}{2}$&$
      {\phi}^{\tau\sigma}_{\psi1}$&$\begin{array}{c}\gamma_1\equiv \left(
	f^\dagger_{0,1,\sigma}\;,\;(i\sigma_y f^{}_{0,1})_\sigma
	\right)\\\Gamma_1\equiv 
\left(
	F^\dagger_{0,1,\sigma}\;,\;(i\sigma_y F^{}_{0,1})_\sigma \right) \end{array}$
      \\
      0&$\frac{1}{2}$&$\frac{1}{2}$&$\sigma$&$\frac{1}{2}$ &$
      {\phi}^{\tau\sigma}_{\psi2}$ & $\begin{array}{c}\gamma_2\\ \Gamma_2\end{array}$
	\\
	$\frac{1}{2}$&$\frac{1}{2}$&0&$\1$&$\frac{1}{2}$&
	$\phi^{\tau\tau'}_\Delta$&
	$
\left(\begin{array}{cc}
	  f^\dagger_{0,1}\vec S \vec \sigma \; i\sigma_y  f^\dagger_{0,2} & 
	  -f^\dagger_{0,1}\vec S \vec \sigma   f_{0,2}
	  \\
	  -f_{0,1}\sigma_y\;\vec S \vec \sigma \;\sigma_yf^\dagger _{0,2} 
	  &-f_{0,1}i\sigma_y\;\vec S \vec \sigma f_{0,2}
	\end{array}\right)
	$
	\\
	0&0&0&$\epsilon$&$\frac{1}{2}$&$\phi_{anis}$&
	$
\vec S (f_{0,1}^\dagger \vec \sigma f_{0,1}-f_{0,2}^\dagger \vec \sigma f_{0,2})$
	\\
	0&0&0&$\1$&$\frac{3}{2}$&${\vec {\cal J}}_{-1}{\vec\phi}_{s}$&
	$
        \vec S (f_{0,1}^\dagger \vec \sigma f_{0,1}+f_{0,2}^\dagger \vec \sigma
	f_{0,2})$
	\\ 
	\hhline{========}
  \end{tabular}
\caption{Highest-weight operators and their dimensions $x^{2CK}$
at the 2CK fixed point. Operators are classified 
by the symmetry group
  $\textrm{SU}_{C1}(2)\times\textrm{SU}_{C2}(2)\times\textrm{SU}_S(2)$ and
the scaling operators of the Ising model.
The constants 
  $c_{1}$ and $c_2$ denote the charge spins 
in channels 1 and 2,
  respectively, while $j$ refers to the spin, and $I$ labels the scaling operators of
  the Ising model: $\1,\sigma,\epsilon$. 
Superscripts $\tau,\tau^\prime=\pm$ refer to  the two components of charge
  spinors, 
while $\sigma=\uparrow,\downarrow$ 
label the components of a spin-$\pm\frac{1}{2}$ spinor.
}
\end{table*}

The operator $\phi_\Delta^{\tau\tau'}$ is of special interest, since it is
relevant at the two-channel Kondo fixed point, just like the spin. 
Its susceptibility therefore diverges logarithmically. 
Good candidates for these operators would be
$\sum_{\sigma\sigma'}\epsilon_{\sigma\sigma'} \gamma_1^{\tau\sigma}\gamma_2^{\tau'\sigma'}$, since
these are spin singlet operators that behave as charge 1/2 spinors in both 
channels.  The $\tau=\tau' = +$ component of this 
operator corresponds to the 
superconducting order parameter 
\be
{\cal O}_{SC} \equiv
f^\dagger_{0,1,\uparrow}f^\dagger_{0,2,\downarrow} -
  f^\dagger_{0,1,\downarrow} f^\dagger_{0,2,\uparrow}
\;, 
\ee
while the $+-$ components describe simply a local operator that hybridizes the
channels, $\sim f^\dagger_{0,1,\sigma} f^{}_{0,2,\sigma}$.

Another candidate would be the operator,  
$\sum_{\sigma\sigma'}\epsilon_{\sigma\sigma'} \Gamma_1^{\tau\sigma} \gamma_2^{\tau'\sigma'}$. 
This operator is also a  local singlet, and has charge spins $c_1=c_2=1/2$. 
It contains the following component of the composite superconducting order parameter 
\be
{\cal O}_{SCC}\equiv 
f^\dagger_{0,1}
\vec S \vec \sigma\;
i\sigma_yf^\dagger_{0,2}\;.
\ee
From their transformation properties it is not obvious, 
 which one of the above superconducting order parameters gives the leading
singularity. However, NRG gives a very solid answer and tells us that, 
while the 
 susceptibility of the traditional operator 
does not diverge 
as the temperature or frequency  goes to zero, 
 that of 
the composite order 
parameter 
does.
It is thus this latter operator that can be 
identified as  $\phi_\Delta^{\tau\tau'}$. 
Note that, in case of  electron-hole
symmetry, the composite hybridization operator 
\be
{\cal O}_{mix}\equiv 
f^\dagger_{0,1}
\vec S \vec \sigma\; f_{0,2}\;
\ee
has the same singular susceptibility as ${\cal O}_{SCC}$ since they 
are both components of the same tensor operator. This is, however, not true
any more  away from electron-hole symmetry. Furthermore, 
superconducting correlations are usually more dangerous, since in the Cooper
channel any small attraction would lead to ordering when a regular lattice model
of two-channel Kondo impurities is considered.


The knowledge of the operator content of the two-channel Kondo fixed point enables
us to describe the effects of small magnetic fields and small channel anisotropies 
$(J_1\ne J_2$).   For energies and temperatures below $T_K$, 
the behavior of the model can be described by the slightly perturbed 
two-channel Kondo fixed point
Hamiltonian. For $J_1\approx J_2$
and in  a small magnetic field, $B\ll T_K$, 
this Hamiltonian can be expressed as
\begin{multline}
{\cal H}={\cal H}^\ast_{2CK}+\\
+D_0^{1/2}\;\kappa_0 \;\phi_{anis}+D_0^{1/2}\;{\vec h}_0\;{\vec\phi}_s +
D_0^{-1/2}\;\lambda_0\; {\vec {\cal J}}_{-1}{\vec\phi}_s+\dots\;.
\label{eq:2CKM}
\end{multline}
Here ${\cal H}^\ast_{2CK}$ is the 2CK fixed point Hamiltonian, and $\kappa_0$
is the  dimensionless coupling to the channel anisotropy field, $\phi_{anis}$,
whereas the effective magnetic field, 
$\vec h_0$, couples to the ``spin field'', $\phi_{s}$. Both of them are relevant 
perturbations at the two-channel  Kondo fixed point and they must vanish 
to end up with the two-channel Kondo fixed point at $\omega,T\to 0$. 
The third  coupling, $\lambda_0$, couples to  the leading irrelevant operator  
(see Tab.~\ref{tab:conf_weights}), which dominates the physics when $\kappa=h=0$.
The  energy  cut-off $D_0$ in Eq.~\eqref{eq:2CKM}  is a somewhat arbitrary
scale: it can be though of as 
the energy scale below which the two-channel  Kondo physics emerges, i.e. 
$D_0 \sim T_K$. 
Then the dimensionless couplings $\kappa_0$, $\lambda_0$ and
$h_0$ are approximately related to the couplings of the original Hamiltonian,
Eq.~(\ref{eq:Kondo_nrg}), as 
\bea
\kappa_0&\approx& K_R\equiv 4 \frac{{J}_1-{
    J}_2}{({J}_1+{ J}_2)^2}\;,\label{eq:def_of_KR}\\
h_0&\approx& {B}/{T_K}\;,\\
\lambda_0&\approx& O\left(1\right)\;.
\eea

However, the arbitrary scale $D_0$ in Eq.~\eqref{eq:2CKM} can be changed at the expense of 
changing the couplings: $D_0\to D, \kappa_0\to \kappa(D)$, $h_0\to h(D)$ and
$\lambda_0\to \lambda(D)$ in such a way that 
the physics below $D_0$ remains unchanged. This freedom translates to
scaling equations,  
 whose 
leading terms 
follow from the conformal field theory results, and read 
\bea\label{eq:scaling_kappa}
\frac{d\kappa(D)}{dx}&=&\frac{1}{2}\kappa(D)+\dots\;,\\
\frac{dh(D)}{dx}&=&\frac{1}{2}h(D)+\dots\;,\label{eq:scaling_b}
\\
\frac{d\lambda (D)}{dx}&=&-\frac{1}{2}\lambda(D)+\dots\;,
\label{eq:scaling_lambda}
\eea 
with $x=-\log D$.  
Solving  these equations with the initial conditions, 
$D=D_0\sim T_K$ and  $h=h_0$, $\kappa=\kappa_0$, 
$\lambda = \lambda_0$, we can read out the energy scales at which 
the rescaled couplings become of the order of one,  
\bea
{T^\ast}&\propto &T_K\; \kappa_0^2 \sim T_K  \frac{({J}_1-{
    J}_2)^2}{({J}_1+{ J}_2)^4} \;,
\nn
\\
{T_h}&\propto & T_K\; h_0^2 \sim  {B^2}/{T_K}\;.
\label{eq:scales}
\eea
At these scales the couplings of the relevant
operators are so large that they can no longer be treated as perturbations.
Below $T^*$ the single channel Kondo behavior is recovered in the more
strongly coupled channel,  while $T_h$ 
can be interpreted as the
 scale where the impurity 
spin dynamics is frozen by the external field. 

The prefactors in Eqs.~(\ref{eq:scales}) 
are somewhat arbitrary, and  depend slightly  on the precise definition
one uses to extract these scales. 
In this paper, we shall use 
the spectral function of the composite 
fermion to define the scales $T_K$ and $T^\ast$.
We define $T_K$ to be the energy  at which for $K_R=0$ the spectral 
function of the composite fermion
takes half of its fixed point value (i.e.\ the value assumed 
at $\omega=0$). Whereas $T^\ast$ is the energy at which for $K_R>0$ it takes
$75\%$ of its fixed point value (see Fig.~\ref{fig:rho_F}). 

It is much harder to relate $T_h$ to a physically measurable quantity. 
We defined it simply through the relation,
\be 
T_h \equiv C_h \frac{B^2}{T_K}\;,
\ee
where the constant was chosen to be $C_h\approx60$. This way $T_h$ corresponds
roughly to the energy 
at which the NFL finite size spectrum crosses over to 
the low-frequency FL spectrum. 
%

\section{NRG calculations}\label{sec:nrg}

Prior to discussing the analytic and numerical features of the response functions,
let us devote this section to the short 
description of the 
NRG procedure used.
All results presented in this paper refer to zero temperature. The NRG
calculations were performed with a discretization
parameter $\Lambda=2$. The sum of the dimensionless couplings 
was  ${\tilde J}_1+{\tilde J}_2=0.4$ for each run. 
The NRG data were computed with a so-called flexible DM-NRG program,
\cite{DMNRG_unpublished} which permits the use of an arbitrary number of Abelian and non-Abelian
symmetries (see Tab.\ \ref{tab:symmetry_generators}), and  
incorporates the spectral-sum conserving density matrix NRG (DM-NRG) algorithm.\cite{DMNRG}  
The DM-NRG method  makes it possible to generate spectral functions 
that satisfy  spectral sum rules with machine precision at $T=0$ temperature. 
For calculations with non-zero magnetic field the use of the DM-NRG
method represents a great advantage over conventional NRG
methods,\cite{Costi_00} 
which loose spectral weights and violate spectral sum rules. Conventional
methods also lead to smaller or bigger jumps in the spectral functions 
at $\omega=0$ 
which hinder
the computation of the universal scaling functions provided by the scale
$T_h$.\cite{2ck_cond} The DM-NRG method solves all these problems if a 
sufficient number of multiplets is 
kept.  
On an ordinary desk-top computer, however, we need to use as 
many symmetries as possible to keep the computation time within reasonable limits.

In the present paper, where we study the electron-hole symmetrical case, 
it is possible  to use the symmetry group $\textrm{SU}_{C1}(2)\times
\textrm{SU}_{C2}(2)\times\textrm{SU}_S(2)$ even in case of channel anisotropy. At
these calculations the maximum number of kept multiplets was 750 in each
iteration. This 
corresponds to the diagonalization of $\approx 85$ matrices with matrix sizes
ranging up to $\approx630$, acting on the
vector space of $\approx 9000$ multiplets consisting of $\approx106000$ states.
In the presence of magnetic field we used the symmetry group
$\textrm{SU}_{C1}(2)\times\textrm{SU}_{C2}(2)\times\textrm{U}_S(1)$, and 
retained a maximum of 1350 multiplets in each iteration, that 
corresponds to the diagonalization of $\approx 150$ matrices with matrix sizes
ranging up to $\approx800$ acting on the vector space of 
$\approx18000$ multiplets consisting of $\approx73000$ states.

In the next sections, we shall see how the knowledge of the operator
content of the two-channel Kondo fixed point can help us to understand the  
analytic structure of the various dynamical correlation functions 
obtained by NRG.

\section{Local fermions' spectral functions and susceptibilities}
\label{sec:fermion}

Let us first  analyse the Green's function of the local 
fermion, ${f}^\dagger_{0,\sigma,\alpha}\leftrightarrow {\vec\gamma}_\alpha$.
The composite fermion's (${F}^\dagger_{0,\sigma,\alpha}\leftrightarrow
{\vec\Gamma}_{0,\alpha}$) Green's function has 
already been looked into in detail in an earlier study of ours.\cite{2ck_cond}
We shall therefore not discuss its analytic properties here but use it 
merely as a reference to define the various energy scales in the NRG 
calculations (see Fig.\ \ref{fig:rho_F}). 
Let us note, however, that 
in the large bandwidth limit, $\omega, T_K\ll D_F $, the spectral function of
the composite fermion and that of the local fermion are simply related, 
\be
\varrho_f(\omega) = \frac 1{2D_F} - \frac \pi 4 J^2 \varrho_F(\omega)\;.
\label{eq:connection_f_F}
\ee
Thus, apart from a trivial constant shift and a minus sign, 
the spectral function of the local fermion is that of the composite fermion,
and all features of $\varrho_F$
are also reflected  in $\varrho_f$.

Before we discuss the NRG results, let us examine what predictions we have
for the retarded Green's function of the operator $f^\dagger_{0,\sigma,\alpha}$
from conformal field theory. By looking at its quantum numbers, this operator can be
identified with the operator $\phi_{\psi\alpha}^{+\sigma}$ (see Tab.\
\ref{tab:conf_weights}), i.e.\ 
\be 
 f^\dagger_{0,\sigma,\alpha} = A_f\; \phi_{\psi\alpha}^{+\sigma} + \dots\;,
\label{eq:f_expansion}
\ee
with the prefactor $A_f \propto 1/\sqrt{{D_F}}$. Note that $A_f$ is a complex number, it
does not need to be real. The dots in the equation above indicate the series of
other, less relevant operators and their descendants, which give subleading corrections to the
correlation function of  $f^\dagger_{0,\sigma,\alpha}$. Furthermore, the expansion 
above holds for the {\em long time behavior}. The ``short time part'' of
the correlation function of $f^\dagger_{0,\sigma,\alpha}$ is not captured by
Eq.~\eqref{eq:f_expansion}, and  gives a constant to ${\cal G}_f(\omega)$ of the order of  $\sim 1/{D_F}$. 
Thus, apart from a prefactor $A_f^2$, a constant shift and subleading terms, 
the Green's function of $f^\dagger_{0,\sigma,\alpha}$ is that of the 
field  $\phi_{\psi\alpha}^{+\sigma}$. 
%
As we discuss it 
shortly in Appendix~\ref{app:scaling_prop}, 
the Fourier transform of the Green's function of any operator of dimension
$x=1/2$ is scale invariant around the two-channel Kondo fixed point. 
%
Since  $\phi_{\psi\alpha}^{+\sigma}$ and thus $f^\dagger_{0,\sigma,\alpha}$ 
have a scaling dimension $1/2$ at the 2CK fixed point, 
it follows that the dimensionless retarded Green's function, ${D_F}\;{\cal
  G}_f(\omega)$, is also scale
invariant,\footnote{Throughout this paper we discuss only retarded Green's
  functions. The other Green's functions are related to them by simple
  analytic relations in equilibrium.}
\bea
{D_F}\;{\cal G}_f(\omega,T) &\equiv&  \hat g_f  \left(\frac{\omega}{D},\frac{T}{D},\kappa(D),
h(D),\lambda(D),\dots\right)\;,
\nonumber
\\
D \frac{{\rm d} \hat g_f}{{\rm d}D} & =& 0\;.
\label{eq:scale_inv_f}
\eea

From Eq.~\eqref{eq:scale_inv_f}, we can deduce various important properties. 
Let us first consider the simplest case, $T=0$ and $\kappa=h=0$. 
Then setting the scale $D$ to  $D_0\sim T_K$ we have
\bea
\hat g_f^{\kappa,h,T=0}(\omega) = \hat g_f  
\left(\frac{\omega}{D_0},\lambda_0,\dots\right).
\label{eq:scaling_c}
\eea
Let us now rescale $D \to |\omega| $, and use the
fixed point scaling equation \eqref{eq:scaling_lambda} to obtain $\lambda(D)$,  
\bea
\hat g_f &=&g_{f}\left(\pm 1,
\sqrt{\frac{|\omega|}{D_0}}\;\;
\lambda_0\right)\;.
\eea 
Assuming that this function is analytic in its second argument we obtain
for $|\omega|\ll T_K$
\bea
\hat g_f^{\kappa,h,T=0} (\omega)&=& \hat g_f\Bigl(\frac  \omega {T_K}\Bigr)\nn
&\approx &g_{\pm\; f}  + g_{\pm\; f}^{\prime}
\;\sqrt {\frac {|\omega|} {T_K}}  +
\dots  
\;,
\eea
with  $g_{\pm\; f}$ and $g_{\pm\; f}^{\prime}$ some complex 
expansion coefficients. Here the subscripts $\pm$ refer to the cases 
$\omega>0$ and $\omega<0$, respectively.
As we discussed above, the constants $g_{\pm\; f}$
depend also on the  short time behavior of   ${\cal G}_f(t)$, and 
are not universal in this  sense. These constants are not independent of 
each other. They are related by the constraint that the Green's function 
must be analytic in the upper half-plane. Furthermore, electron-hole 
symmetry implies that $g_{+\; f}= g_{-\; f}$
and $g_{+\; f}^\prime= - (g_{-\; f}^\prime)^*$.
 
Relations similar to the ones above hold for the dimensionless spectral
function. 
It is defined as 
\be 
\hat \varrho_f(\omega) \equiv -\frac 1 \pi \;\im \hat g_f (\omega)\;, 
\ee
and  
assumes the following simpler form  at 
small frequencies in case of electron-hole symmetry,  
\be
\hat \varrho_f^{T,\kappa,h=0}(\omega) = {r_f}
+ r_f^\prime  
\;\;
\sqrt{\frac {|\omega|}{T_K}} 
+
\dots  
\;.
\ee

\begin{figure}[t]
  \includegraphics[width=0.9\columnwidth,clip]{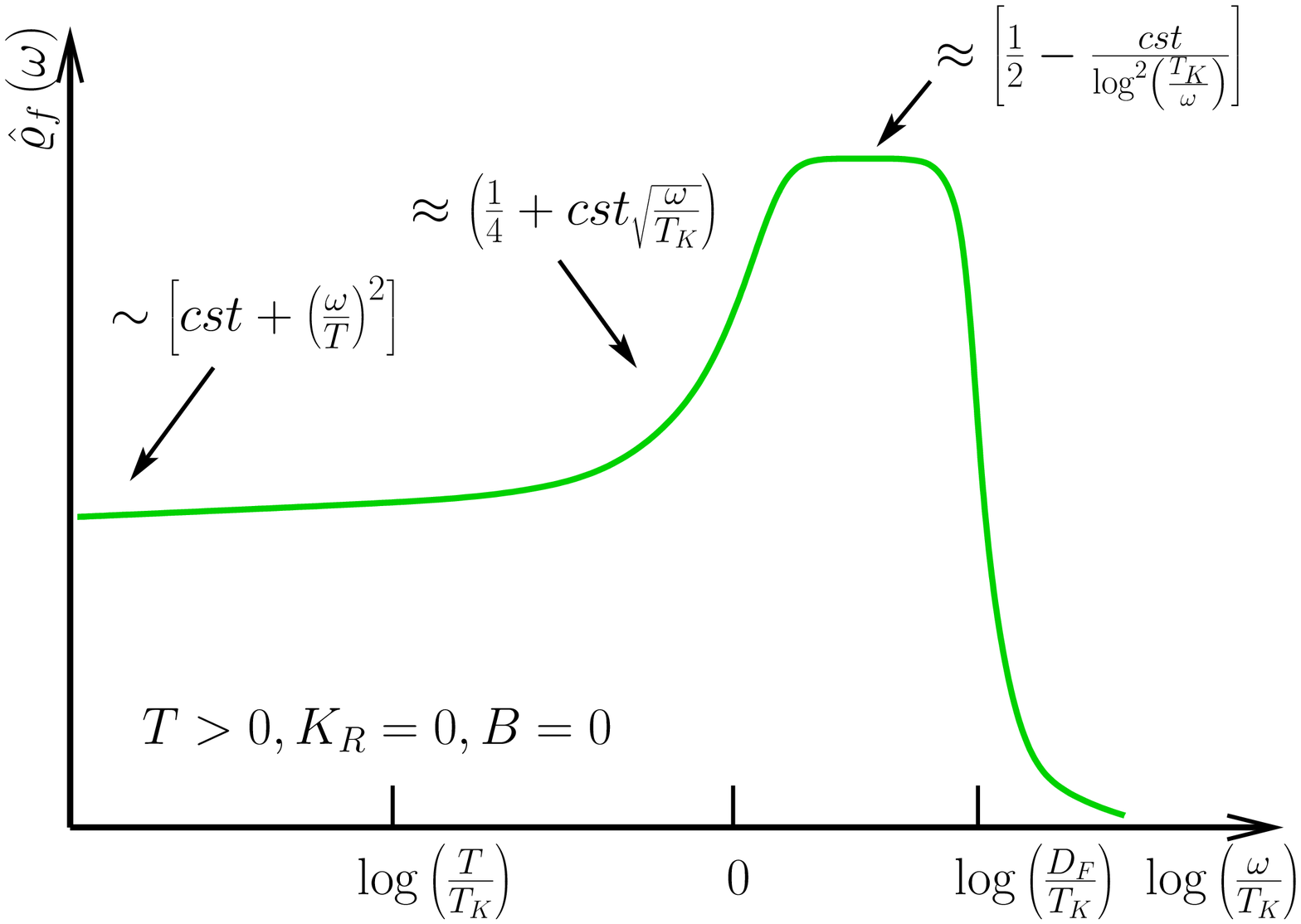}
  \includegraphics[width=0.9\columnwidth,clip]{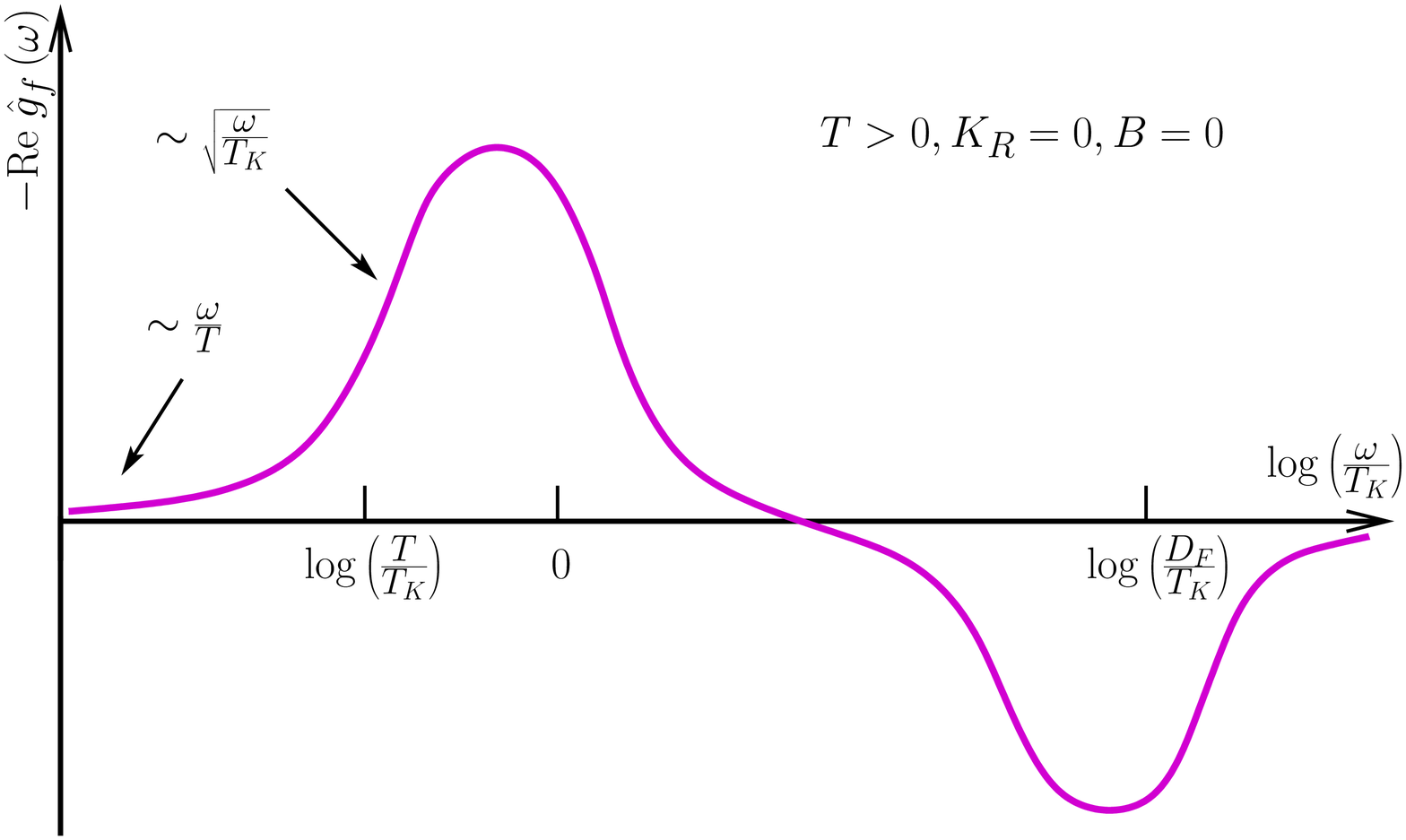}
  \caption{(color online) 
    (top) Sketch of the dimensionless spectral
    function $\hat{\varrho}_f=D_F\varrho_f$  of $f^\dagger_{0,1,\sigma}$, and
    (bottom) the real part of its dimensionless Green's function,
    $\re\hat{g}_f= D_F\; \re{\cal G}_f$
    for $T>0$ and $K_R=0,B=0$  as a function of $\log\left(\omega/T_K\right)$.
    Asymptotics indicated for $\omega<T_K$ were derived through scaling
    arguments. The large $\omega$-behavior is a result of perturbation theory.
  }
  \label{fig:rho_f_finT_sketch}
\end{figure}

\begin{figure}[t]
  \includegraphics[width=0.9\columnwidth,clip]{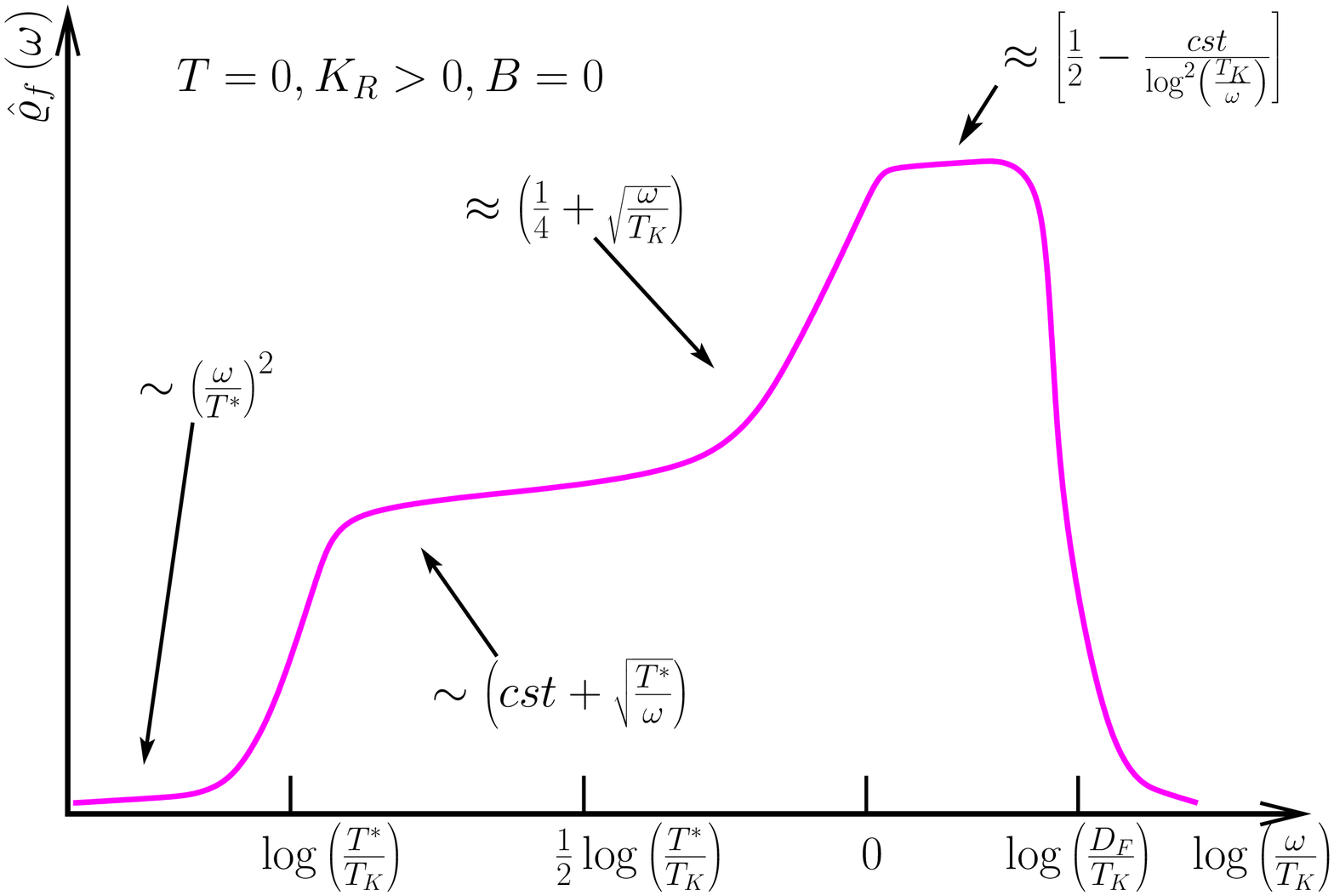}
  \includegraphics[width=0.9\columnwidth,clip]{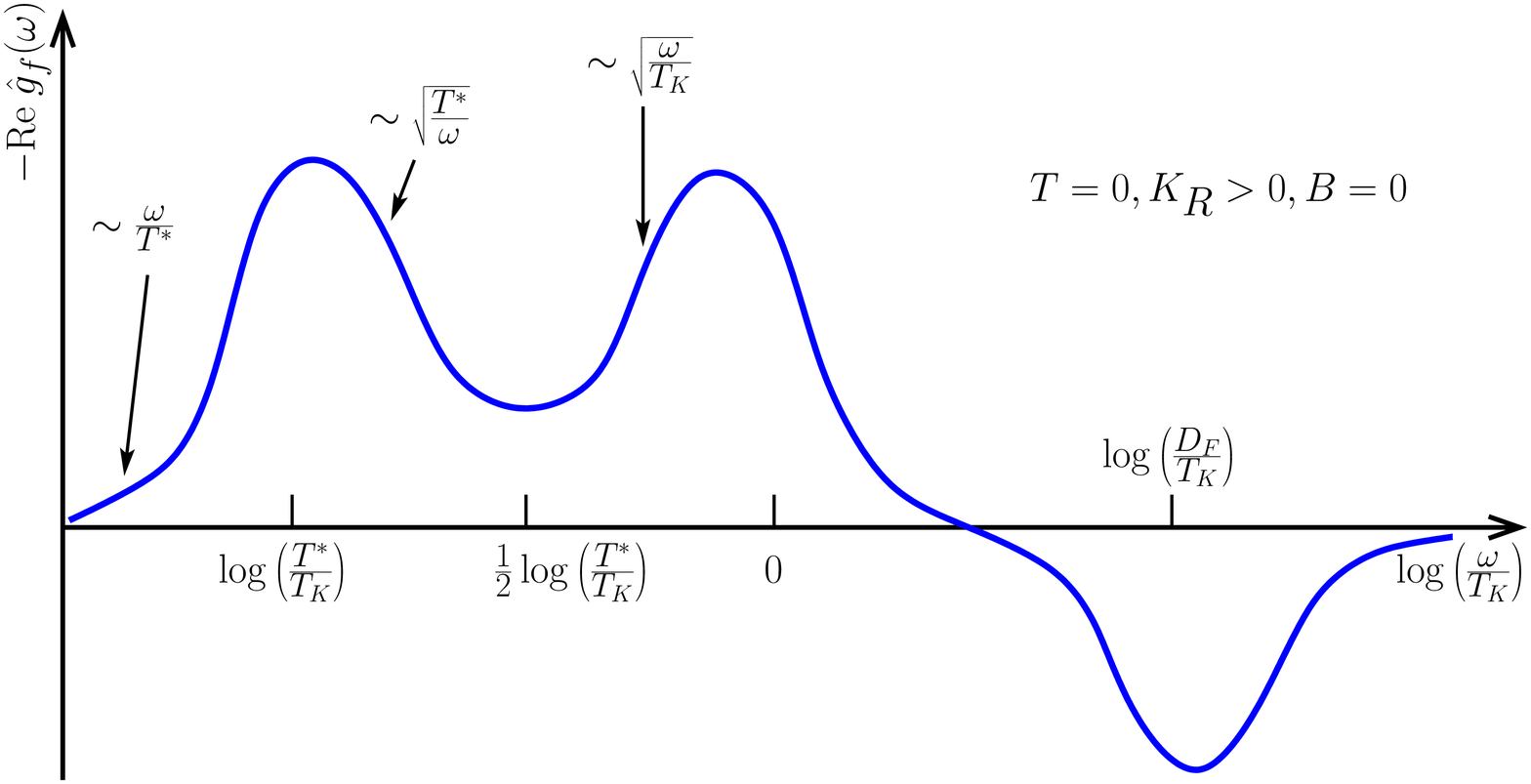}
  \caption{(color online) (top) 
    Sketch of the dimensionless spectral function of $f^\dagger_{0,1,\sigma}$:
    $\hat{\varrho}_f=D_F\;{\varrho}_f$, and (bottom) the real part of its 
    dimensionless Green's function: $\re\hat{g}_f=D_F\;\re{\G}_f$ 
    for $T=0,K_R>0$ and $B=0$  as a function of
    $\log\left(\omega/T_K\right)$. 
    Asymptotics indicated for $\omega<T_K$ were derived through scaling
    arguments. The large $\omega$-behavior is a result of perturbation theory.
  }
  \label{fig:rho_f_finT*_sketch}
\end{figure}

For $\omega\gg T_K$ the scaling dimension of the local fermion is
$x^{free}_f=1/2$ corresponding to an $\omega$-independent spectral function. 
Perturbation theory in $J$ amounts to logarithmic corrections of the form:
 $1/2-cst/\log^2(T_K/\omega)$,
 as it is sketched in the upper parts of 
Fig.-s \ref{fig:rho_f_finT_sketch} and \ref{fig:rho_f_finT*_sketch}.

For  $T\ne 0$, and  $\kappa=h=0$ using  
similar arguments as before, but now rescaling $D \to T $ we find 
\bea
\hat g_f^{\kappa,h=0}(\omega) &=& 
\hat g_f\left( \frac \omega T, \frac  T {T_K},\lambda_0\right)\nn
 &\equiv&
\hat g_f\left(\frac{\omega}{T},1, 
\sqrt{\frac{T}{D_0}}\;\;  
\lambda_0,\dots\right).
\eea
Then by expanding $\hat g_f$ 
we obtain the following scaling form for the
low temperature behavior of the spectral function, 
\bea
\hat \varrho_f^{h,\kappa=0}(\omega) &=&
\Theta_{f}\left(\frac{\omega}{T}\right)+
\sqrt{{\frac{T}{T_K}}}
\;\;
{\tilde\Theta}_{f}\left(\frac{\omega}{T}\right)+\dots\;,
\label{eq:rho_f}
\eea 
with $ \Theta_f$ and  $\tilde  \Theta_f$ universal scaling functions. Note that we
made no assumption on the ratio $\omega/T$, but both $\omega$ and $T$ 
 must be smaller
than $T_K$. 
The asymptotic properties of $ \Theta_{f}$ and ${\tilde  \Theta}_{f}$ can be extracted by making
use of the facts that $(i)$ $\hat g_f(\omega,T)$ must be analytic for $\omega\ll
T$,   $(ii)$ that  Eq.~\eqref{eq:rho_f}  should reproduce the
 $T\to 0$ results in the 
limit $\omega\gg T$, and $(iii)$ that by electron-hole symmetry, 
$\hat \varrho_f$ must be an even function of $\omega$. 
The issuing asymptotic properties together with those of the other scaling functions
defined later are summarized in Table~\ref{tab:asymp}.
The asymptotic properties of the real part, ${\rm Re}\;\hat g_f$, 
can be extracted from those of  
$\hat \varrho_f$ by performing a Hilbert transform
\bea
 {\rm
   Re}\;\hat g_f(\omega)= \mathcal{P}\ds{\int} {{\rm d}{\tilde\omega}}\;
 \frac{\hat \varrho_f({\tilde\omega})}
{{\omega}-\tilde \omega}
\eea
with $\mathcal{P}$ the principal part.  The obtained  features are sketched 
in Fig.~\ref{fig:rho_f_finT_sketch} for $T>0$ and $\kappa=h=0$.


Let us now investigate the effect of channel anisotropy, i.e.\ $\kappa\neq0$ 
at $T=0$ temperature and no magnetic field $h=0$. 
In this case, we can rescale  $D$  to $D = |\omega|$ to obtain
\bea 
\hat \varrho_f
^{T,h=0}({\omega})= 
\K^{\pm}_{f}\left(\frac{\omega}{T^*}\right)
+
\sqrt{\frac{|\omega|}{T_K}}
\;\;{\tilde
\K}^\pm_{f}\left(\frac{\omega}{T^*}\right) + \dots \;,\nn
\eea 
with $T^*$ the anisotropy scale defined earlier. 
The superscripts $\pm$ refer to the cases of 
positive or negative anisotropies: the superscript 
``$+$'' is used   when the coupling is larger in the channel where we measure the 
Green's function of $f^\dagger_{0,\alpha,\sigma}$. 
The asymptotics of the universal functions
$ \K^{\pm}_{f}$ and ${\tilde  \K}^\pm_{f}$ can be obtained through similar
scaling arguments as before and they differ only slightly
from those of $ \Theta_{f}$ and ${\tilde \Theta}_{f}$ (see
Table~\ref{tab:asymp} for a summary).
The  properties of 
$\hat \varrho_f^{T,h=0}({\omega})$
are summarized in Fig.~\ref{fig:rho_f_finT*_sketch}. 
A remarkable feature of the spectral function is that it contains
a correction $\sim \sqrt{T^*/|\omega|}$. This correction 
can be obtained by doing perturbation theory in the small parameter  
$\kappa(\omega)$ at the two-channel Kondo fixed point.

\begin{table*}[htb]\label{tab:asymp}
  \begin{tabular}{cccc}
    \hhline{====}
    Scaling Function&$\begin{array}{c}\textrm{Asymptotic
	Form}\\\begin{array}{cc}x\ll1\;,\quad\quad\quad&\quad1\ll
      x\end{array}\end{array}$&$\begin{array}{c}\textrm{Scaling
	Variable}\\x\end{array}$
      &$\begin{array}{c}\textrm{2CK}\\\textrm{Scaling Regime}\end{array}$\\
    \hline
    $\begin{array}{c}{\Theta}_f\;(x)\\{\tilde\Theta}_f\;(x)\end{array}$
    &$\begin{array}{cc}\theta_{\;f}^{\;0}+\theta_{\;f}^{\;0\;\prime}\;x^2\;,&\quad\theta_{f}^{\infty}\\
      {\tilde\theta}_{\;f}^{\;0}+{\tilde\theta}_{\;f}^{\;0\;\prime}\;x^2\;,&\quad{\tilde\theta}_{f}^{\infty}\;x^{1/2}\end{array}$&$\omega/T$&$T\lessapprox\omega$\\
    $\begin{array}{c}\K^\pm_f\;(x)\\{\tilde\K}^\pm_f\;(x)\end{array}$&
    $\begin{array}{cc}\kappa^{\pm}_{f,0}+\kappa^{\pm\;\prime}_{f,0}\;
    x^2\;,&\quad
    \kappa^\pm_{f,\infty}+\kappa^{\pm\;\prime}_{f,\infty}\;\left|\frac{1}{x}\right|^{1/2}\\
    {\tilde\kappa}^\pm_{f,0}\;
    \left|x\right|^{3/2}\;,&\quad{\tilde\kappa}^\pm_{f,\infty}\end{array}$&$\omega/T^*$&$T^{**}_f\lessapprox\omega\;,
    \quad T^{**}_f\propto\sqrt{T^*T_K}$\\
    $\begin{array}{c}\B_{f,\sigma}\;(x)\\{\tilde\B}_{f,\sigma}\;(x)\end{array}$
    &$\begin{array}{cc}\beta^{\;0}_{{f,\sigma}}+\beta^{\;0\;\prime}_{{f,\sigma}}\;x^2\;,\quad
    &\beta_{f,\sigma}^{\;\infty}+\beta^{\;\infty\;\prime}_{f,\sigma}\;\left|\frac{1}{x}\right|^{1/2}\\
      {\tilde\beta}_{f,\sigma}^{\;0}\;\left|x\right|^{3/2}\;,
    &\quad{\tilde\beta}_{f,\sigma}^{\;\infty}
    \end{array}$&$\omega/T_h$&$T^{**}_h\lessapprox\omega\;,\quad T^{**}_h\propto\sqrt{T_hT_K}$\\
    $\begin{array}{c}{\Theta}_S\;(x)\\{\tilde\Theta}_S\;(x)\end{array}$&$\begin{array}{cc}\theta^{\;0}_{S}\;
    x\;,&\quad\theta_{S}^{\infty}\;\sgnx\\
    {\tilde\theta}^{\;0}_{{ S}}\;x\;,
    &\quad{\tilde\theta}_{{ S}}^\infty\;{\rm sgn}(x)\; \left|x\right|^{1/2}
    \end{array}$&$\omega/T$&$T\lessapprox\omega$\\
    $\begin{array}{c}\K_{S}\;\left(x\right)\\{\tilde\K}_{S}\;\left(x\right)\end{array}$&
    $\begin{array}{cc}\kappa_S^{\;0}\; x\;,
      &\quad{\kappa_{{S}}^\infty}\;\sgnx +
      \kappa_{{S}}^{\infty\;\prime\prime}\; \frac{1}{x}\\
	  {\tilde\kappa}_S^{\;0}\;\sgnx \left|x\right|^{1/2}\;,
	  &\quad {\tilde\kappa}_S^\infty\;\sgnx
    \end{array}$&$\omega/T^*$&$T^{**}_s\lessapprox\omega\;,\quad T^{**}_s\propto\left({T^*}^2T_K\right)^{1/3}$\\
    $\begin{array}{c}\B_{S,z}\;\left(x\right)\\{\tilde\B}_{S,z}\left(x\right)\end{array}$
    &$\begin{array}{cc}\beta_{S,z}^{\;0} \; x\;,
      &\quad\beta_{S,z}^\infty  + \beta_{S,z}^{\infty\;\prime}\;
      \left|\frac{1}{x}\right|^{1/2}\\
	   {\tilde\beta}_{S,z}^{\;0}\; \left|x\right|^{1/2}\;,
	   &\quad{\tilde\beta}_{S,z}^{\infty}
    \end{array}$&$\omega/T_h$&$T^{**}_h\lessapprox\omega\;,\quad T^{**}_h\propto\sqrt{T_hT_K}$\\
    \hhline{====}
    \end{tabular}
\caption{Asymptotic behavior of the universal cross-over functions. 
  At finite temperature, the boundary of the two-channel Kondo scaling 
  regime is set by the temperature. At zero temperature, the various 
  boundaries of the 2CK scaling 
  regime derive from the competition between the leading
  irrelevant operator and the relevant perturbation.
}
\end{table*}

From the asymptotic forms in  Table~\ref{tab:asymp}
we find that in the local fermion's susceptibility 
a new scale, $T^{**}_f\sim\sqrt{T^\ast T_K}$ appears as a result of the
competition between the leading irrelevant operator and the channel
anisotropy:\cite{2ck_cond} It is only in the regime $T^{**}_f< \omega < T_K$ that the leading 
irrelevant operator determines the dominant scaling behavior of the local
fermion's susceptibility, i.e., we
observe the true two-channel Kondo physics. 
The expected properties of $\hat \rho_f$ and the real part of its
dimensionless Green's function $\hat g_f$ in the presence of channel asymmetry
are summarized in Fig.~\ref{fig:rho_f_finT*_sketch}.
These analytic expectations are indeed met by our NRG results. 

\begin{figure}[b]
  \includegraphics[width=0.9\columnwidth,clip]{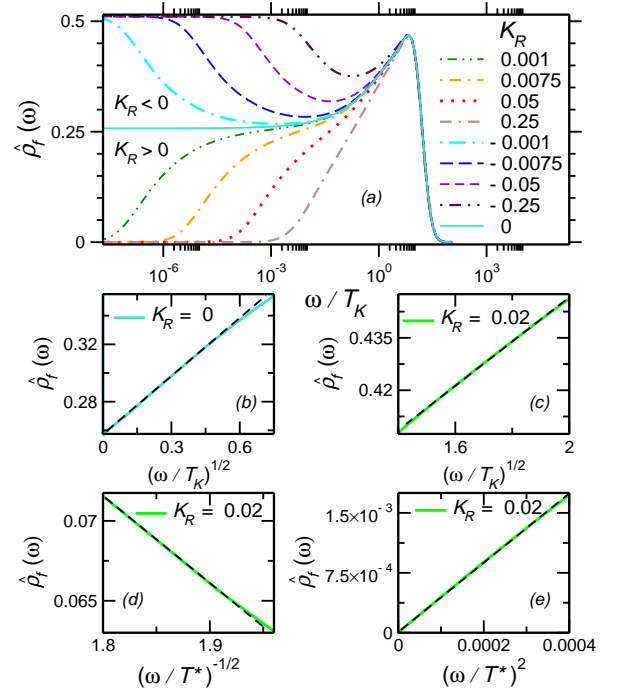}
  \caption{(color online) $(a)$ Dimensionless spectral function of $f_{0,1,\sigma}$:
    $\hat\varrho_f(\omega)=D_F\;\varrho_f(\omega)$ as a function of $\omega/T_K$ 
    for different values of $K_R$. $(b-e)$ 
    Numerical confirmations of the low-frequency asymptotics derived through 
    scaling arguments in Sec.\ \ref{sec:fermion}. 
    Dashed straight lines are to demonstrate 
    deviations from the expected $\sqrt{\omega}$-like $(b-c)$, 
    $1/\sqrt{\omega}$-like $(d)$ and $\omega^2$-like $(e)$
    behavior. In plots $(c-e)$ $T^*/T_K=2.4\times 10^{-4}$. 
  }\label{fig:rho_f_chani_TK}
\end{figure}


Fig.\ \ref{fig:rho_f_chani_TK}.$(a)$ depicts the spectral function of
$f^\dagger_{0,1,\sigma}$ for several values of $K_R$ as a function of
$\omega/T_K$ on a logarithmic scale. The overall scaling is very similar to
the one sketched in Fig.~\ref{fig:rho_f_finT*_sketch}, except that the high
temperature plateau is missing; this 
 is due to the relatively large value of
$T_K$, which is only one decade smaller than the bandwidth cut-off. 
Figures \ref{fig:rho_f_chani_TK}.$(b-e)$ are the numerical confirmations 
of the asymptotics stated. In all these figures
dashed straight lines are to demonstrate deviations from the expected
behavior.
In Fig.\ \ref{fig:rho_f_chani_TK}.$(b)$ we show the 
square 
root-like asymptotics in the 2CK scaling regime for the channel
symmetric case. This behavior is a consequence of the dimension of the leading
irrelevant operator as it has just been discussed. 
In Fig.\ \ref{fig:rho_f_chani_TK}.$(c)$ the same asymptotics is shown 
in the same region in case of a finite channel anisotropy, whereas below
them Fig.\ \ref{fig:rho_f_chani_TK}.$(d)$ demonstrates 
$\left(1/\omega\right)^{1/2}$-like behavior resulting from the
relevant perturbation of the 2CK fixed point Hamiltonian with channel
anisotropy. In Fig.\ \ref{fig:rho_f_chani_TK}.$(e)$ the FL-like 
$\omega^2$-behavior is recovered below $T^*$, which is typical of 
fermionic operators in the 1CK scaling regimes.

\begin{figure}[t]
  \includegraphics[width=0.9\columnwidth,clip]{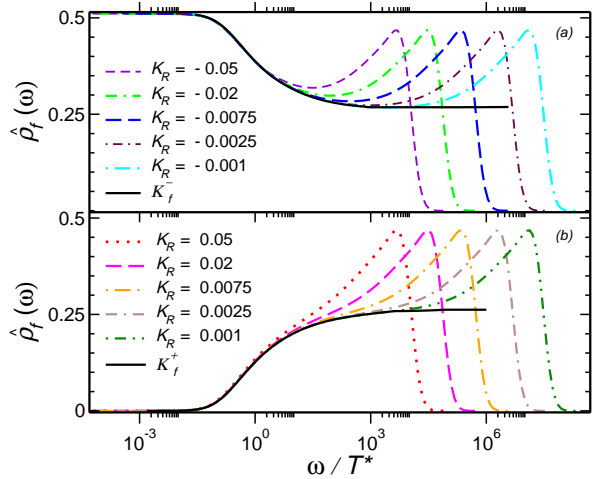}
  \caption{(color online) Universal collapse of the dimensionless 
    spectral functions, $\hat \varrho_f=D_F\;\varrho_f$
    (with $f$ in channel 1) to two scaling curves, $\K^\pm_f$ as a function of
    $\omega/T^\ast$ for positive $(b)$ and negative 
    $(a)$ values of $K_R$.}
  \label{fig:rho_f_chani_Tstar}
\end{figure}

In Figs.\ \ref{fig:rho_f_chani_Tstar}.$(a-b)$ we show the universal scaling curves, 
$\K^\pm$
that connect the two-channel and single channel fixed points
at low-frequencies as a function of $\omega/T^\ast$. 
They were computed  from runs with negative and positive
values of $K_R$. This universal behavior is violated 
for 
values of $K_R$  higher
than the highest ones shown in Fig.\ \ref{fig:rho_f_chani_Tstar},
where $T^*$ becomes comparable to $T_K$. 

\begin{figure}[b]
  \includegraphics[width=0.9\columnwidth,clip]{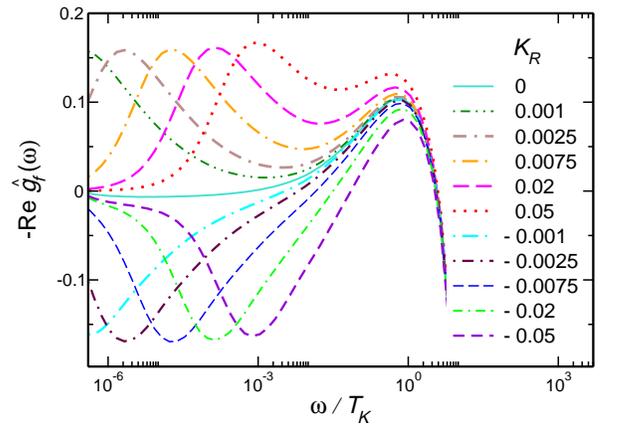}
  \caption{(color online) Real part of the dimensionless Green's function:
    $\re\hat g_f=D_F\;\re\G_f$ (with $f^\dagger$ in channel 1) as a function of $\omega/T_K$ 
    for different values of $K_R$. From among 
    the three peaks sketched in Fig.\ \ref{fig:rho_f_finT*_sketch} only the two peaks 
    around $T^*$ and $T_K$ are shown.
  }
  \label{fig:re_g_f_chani_TK}
\end{figure}
The real parts 
of the local fermion susceptibilities are plotted in 
 Fig.\ \ref{fig:re_g_f_chani_TK} for several values of $K_R$. They were 
obtained by performing the Hilbert transformations numerically. They should show a three-peak
structure based on the analytic considerations (see 
Fig.\ \ref{fig:rho_f_finT*_sketch}). There are two   
low-frequency peaks clearly visible, associated with the cross-overs at $T^*$
and $T_K$. Furthermore there should be a non-universal peak at the cut-off.
For relatively large channel
anisotropies, where $T^*\sim T_K$, the former two peaks cannot be clearly separated 
in Fig.\ \ref{fig:re_g_f_chani_TK}. 
Also,  due to the large value of $T_K\sim$ the band cut-off, $D_F$, the peak at $\omega\sim T_K$
and the smeared singularity at $\omega=D_F$ merge to a single non-universal
feature in our NRG curves.

\begin{figure}[t]
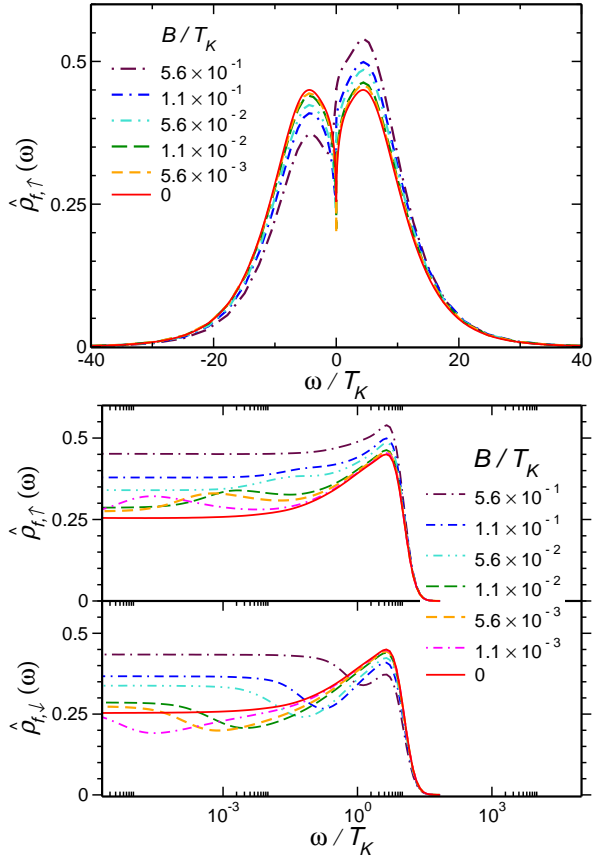

  \includegraphics[width=0.9\columnwidth,clip]{fig_6_RR/rho_f_Bneq0_lin_RR}
  \includegraphics[width=0.9\columnwidth,clip]{fig_6_RR/rho_f_Bneq0_log_RR}
  \caption{(color online) Top: Dimensionless spectral function of
    ${f}_{0,1,\uparrow}:$ ${\hat\varrho}_{f,\uparrow}=D_F\;\varrho_{f,\uparrow}$ 
    for different values of $B$ as a 
    function of $\omega/T_K$ on linear scale.
    Bottom: Dimensionless spectral function of ${f}_{0,1,\uparrow}$ $(a)$ and of 
    ${f}_{0,1,\downarrow}$ $(b)$ for different values of $B$ as a 
    function of $\omega/T_K$ on logarithmic scale. 
  }
  \label{fig:rho_f_Bneq0_lin_log}
\end{figure}

Let us now turn to  the effect of a finite magnetic field, 
$B\neq0$ for the case $T=0,K_R=0$.  
As $h$ and $\kappa$ scale the same way in the 2CK scaling regime, the
argument concerning the $\kappa\neq0$ case can be repeated with minor modifications. 
Now, however, the spin $\textrm{SU}_S(2)$ symmetry is violated, 
and therefore the spectral functions of $f^\dagger_{0,\alpha,\uparrow}$
and $f^\dagger_{0,\alpha,\downarrow}$ become different, and they are no longer
even either.
Nevertheless, due to particle-hole
symmetry, they 
are still related through the relations
\bea 
\hat \rho_{f,\uparrow}\left(\omega,T,\kappa,h,\dots\right)  
&=&\hat \rho_{f,\downarrow} \left(-\omega,T,\kappa,h,\dots\right)\;,
\nn
\hat \rho_{f,\uparrow}\left(\omega,T,\kappa,h,\dots\right) &=& 
\hat \rho_{f,\downarrow} \left(\omega,T,\kappa,-h,\dots\right)\;.
\eea 
We are thus free to choose the orientation of the magnetic field downwards. 
Then, after rescaling $D\to |\omega|$  we get
\begin{multline}
\hat \varrho_{f,\sigma}^{\kappa,T=0}(\omega)
=\B_{f,\sigma}\left(\frac{\omega}{T_h}\right)+
\sqrt{\frac{|\omega|}{T_K}}\;\;
{\tilde\B}_{f,\sigma}\left(\frac{\omega}{T_h}\right)+\dots\;,
\end{multline}
where the label $\sigma$ refers to the different spin components and
$\B_{f,\sigma}$ and ${\tilde\B}_{f, \sigma}$ are yet another pair of
universal cross-over functions. The asymptotic properties of the 
functions $\B_{f,\sigma}$ and $\tilde \B_{f,\sigma}$
are summarized in Table~\ref{tab:asymp}.

\begin{figure}[t]
  \includegraphics[width=0.9\columnwidth,clip]{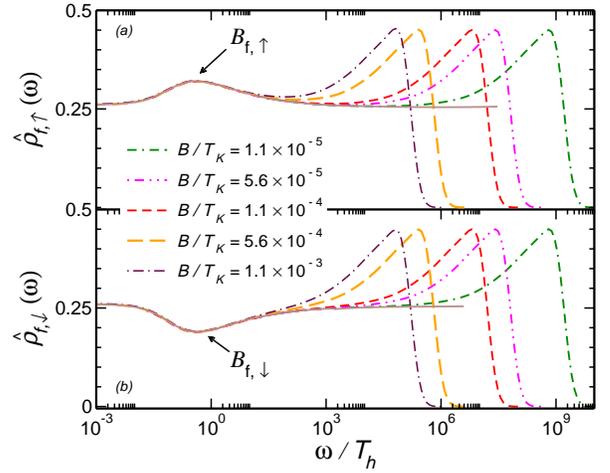}
  \caption{(color online) Universal collapse of the dimensionless spectral functions: $\hat
    \varrho_{f,\uparrow}=D_F\;\varrho_{f,\uparrow}$ and $\hat\varrho_{f,\downarrow}=D_F\;\varrho_{f,\downarrow}$ 
    to two scaling curves: $\B_{f,\uparrow}$ and $\B_{f,\downarrow}$
    for sufficiently small, non-zero values of $B$ as a function of $\omega/T_h$.
  }
  \label{fig:f0_Bneq0_double_scaled}
\end{figure}

\begin{figure}[b]
  \includegraphics[width=0.9\columnwidth,clip]{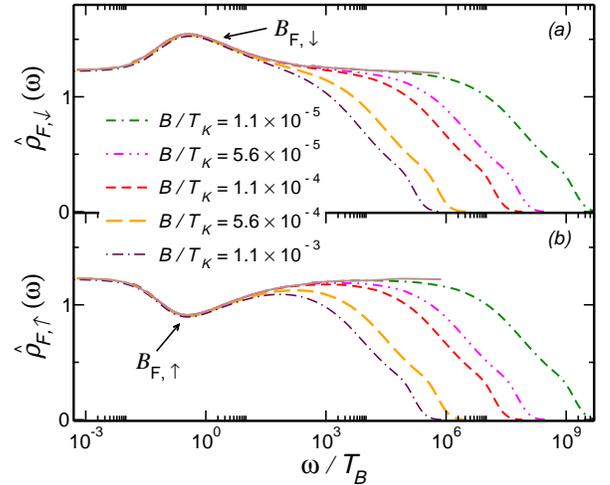}
  \caption{(color online) 
    Universal collapse of the dimensionless spectral functions of the
    composite fermion operator: $\hat
    \varrho_{F,\uparrow}=D_F\;\varrho_{f,\uparrow}$ and $\hat\varrho_{F,\downarrow}=D_F\;\varrho_{F,\downarrow}$ 
    to two scaling curves: $\B_{F,\uparrow}$ and $\B_{F,\downarrow}$
    for sufficiently small, non-zero values of $B$ as a function of $\omega/T_h$.
  }
  \label{fig:F_in_h_scaled}
\end{figure}

Fig.\ \ref{fig:rho_f_Bneq0_lin_log} shows the spectral functions $\hat
\varrho_{f,\sigma}$ 
as a function of $\omega/T_K$ on linear and
logarithmic scales for different magnetic field values. 
The same curves are depicted  as a function of 
$\omega/T_h$ in Fig.\ \ref{fig:f0_Bneq0_double_scaled},
which demonstrates the existence of the universal scaling curves, 
$\B_{f,\sigma}$, i.e.\ that by using the scale, 
$T_h$ the local fermion's spectral functions can be scaled on top of each
other for small enough magnetic fields. In this magnetic field region, we find a
peak at $T_h$ for the spin-$\uparrow$ component of $f^\dagger$, while at the same
place there is a dip 
for the spin-$\downarrow$ component. 
This is a remarkable feature that is associated with inelastic scattering 
off the slightly polarized impurity spin. In fact, the same uinversal 
features also  appear in the spectral functions of the composite fermions,
which we compute independently and which are directly
related to those of the conduction electrons by Eqn.~\eqref{eq:connection_f_F}.\cite{Costi_00}
The rescaled spectral functions $\hat\varrho_{F,\sigma}(\omega)$ are shown in 
Fig.\ref{fig:F_in_h_scaled}. 

\begin{figure}[t]
  \includegraphics[width=0.9\columnwidth,clip]{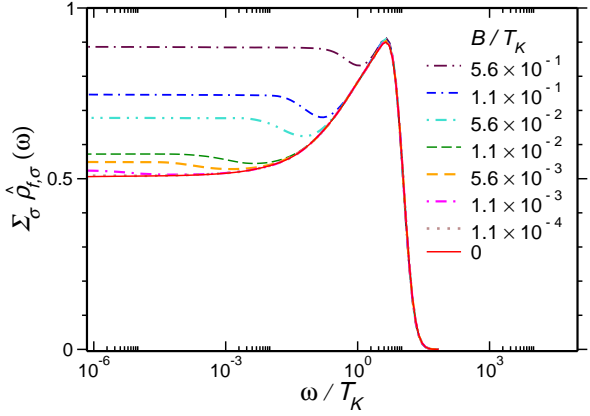}
  \caption{(color online) Sum of the dimensionless spectral functions: ${\hat \varrho}_{f,\uparrow}=D_F\;{\varrho}_{f,\uparrow}$ and
    ${\hat\varrho}_{f,\downarrow}=D_F\;{\varrho}_{f,\uparrow}$ for different values of $B$ as a function of $\omega/T_K$.
  }
  \label{fig:rho_f_sigma_sum_Bneq0}
\end{figure}

Although this numerical evidence can be obtained by conventional
NRG methods not using the density matrix, this is no longer true for the sum of the
local fermions spectral function over the different spin components. In fact,
for this quantity universal scaling curves in the presence of magnetic field cannot be
obtained using NRG because of the increase in the size of the numerical errors at
low-frequencies and the mismatch between the positive and negative frequency
parts of the spectral functions. The sum of the local fermion's 
spectral function over the two spin components is depicted in Fig.\ \ref{fig:rho_f_sigma_sum_Bneq0}
as a function of $\omega/T_K$. 
Here the splitting of the Kondo resonance in the energy-dependent scattering cross section 
appears as a minimum at $\omega\sim
T_h$. Unfortunately, for even smaller magnetic fields the accuracy 
of our numerical data 
is insufficient to tell if the  splitting of the Kondo resonance survives in the
limit $B\to 0$, as conjectured in Ref.~\onlinecite{2ck_cond}. In the data with 
$B/T_K > 1.1 \times 10^{-4}$, there seems to be always a shallow minimum 
in the spectral function, and we see no indication for  crossing of the curves 
as the 
 magnitude of 
the field is reduced. If there is indeed no crossing of the spectral functions 
and if the deviation from the $\sqrt{|\omega|}$-behavior
indeed starts at $\omega\approx  T_h\sim B^2/T_K$, which is the only natural
assumption, then, from exact  Bethe Ansatz results it would immediately 
follow that there must always be a splitting of the Kondo resonance, since 
$\sum_\sigma[\varrho_{f\sigma}(\omega=0,B)-\varrho_{f\sigma}(\omega=0,0)] \sim
B \ln(T_K/B)$,\cite{Pustilnik_04} while 
$\sum_\sigma[\varrho_{f\sigma}(\omega=T_h,B)-\varrho_{f\sigma}(\omega=0,0)] \sim
|B|$ would follow from the pure $\sqrt{|\omega|}$-dependence of the spectral
function at $B=0$. However, these analytical arguments do not constitute a 
real proof.  
  
%
 
With small modifications, the analysis presented in this subsection carries over to 
essentially any fermionic operator that has quantum numbers 
$c_1=j=1/2$ or $c_2=j=1/2$ and has a finite overlap with 
the primary fields $\phi_{\psi 1}$ and $\phi_{\psi 2}$, only the 
high-frequency behavior ($\omega>T_K$) and the normalization factors become different.
Typically, a local 
operator having the same charge and spin quantum numbers as  $\phi_{\psi\alpha}$
will have a finite overlap with them. However, in some cases the internal 
Ising quantum number of an operator may prevent an overlap and,  
of course, one can also construct operators by, say, differentiating with
respect to the time, that would correspond to descendant fields.

\section{Spin spectral functions and susceptibilities}\label{sec:spin}

In this section, we shall discuss the properties of the spin operator,
$\vec S$, which is the 
 most obvious
example of a bosonic operator of spin $j=1$ and
charge quantum numbers $c_1=c_2=0$ that overlaps 
 with the scaling operator $\phi_s$. There 
 are, however, 
many operators that have the same quantum numbers: two examples are the 
so-called channel spin operator, 
\be
{\vec S}_{C}\equiv f_{0,1}^\dagger \vec\sigma f_{0,1}
-f_{0,2}^\dagger \vec\sigma f_{0,2}\;,
\ee
or a composite channel spin operator 
\be
{\vec S}_{CC} 
\equiv F_{0,1}^\dagger \vec\sigma f_{0,1}
-F_{0,2}^\dagger \vec\sigma f_{0,2}\;.
\ee
Our discussion
can be easily generalized to these operators with minor modifications.

The analysis of the spin spectral function goes along the lines 
of the previous subsection. First we recall 
that the field $\vec \phi_s$ appears in the expansion of the spin operator,  
\be\label{eq:spin_coeff} 
\vec S = A_s\; \vec \phi_s +\dots\;,
\ee
with  $A_s \sim 1/\sqrt{T_K}\sim 1/\sqrt{D_0}$. Therefore, the appropriate 
dimensionless scale invariant Green's function (usually referred to as the
{\it dynamical spin susceptibility}) 
is defined as 
\begin{multline}
{\hat g}_{S}
\left(\frac{\omega}{D},\frac{T}{D},\kappa_0,h_0,\dots\right)  \equiv 
T_K \;G_{S}\left(\omega,T,\kappa_0 ,\dots,D_0\right)\;.
\end{multline}

\begin{figure}[t]
  \includegraphics[width=0.9\columnwidth,clip]{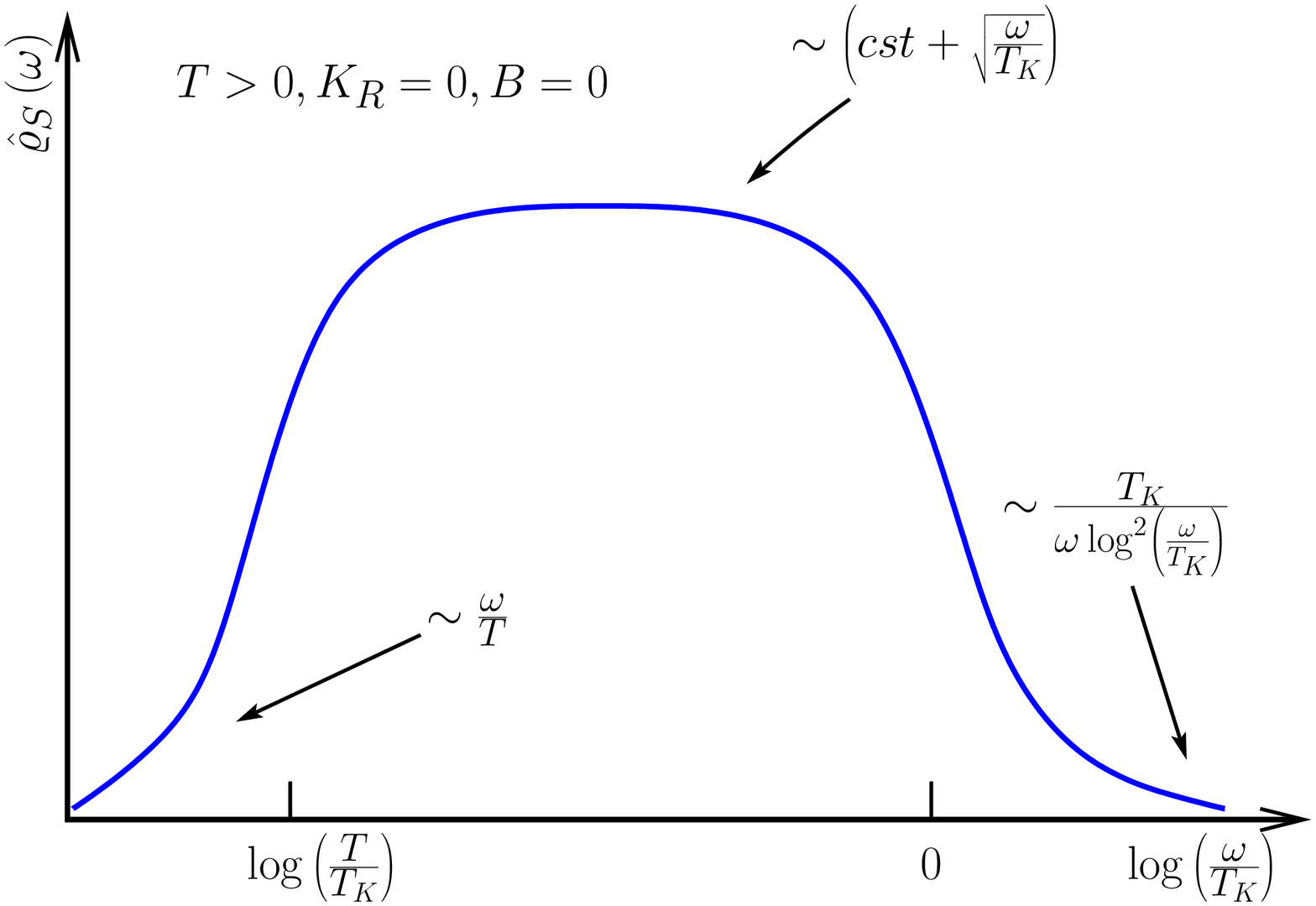}
  \includegraphics[width=0.9\columnwidth,clip]{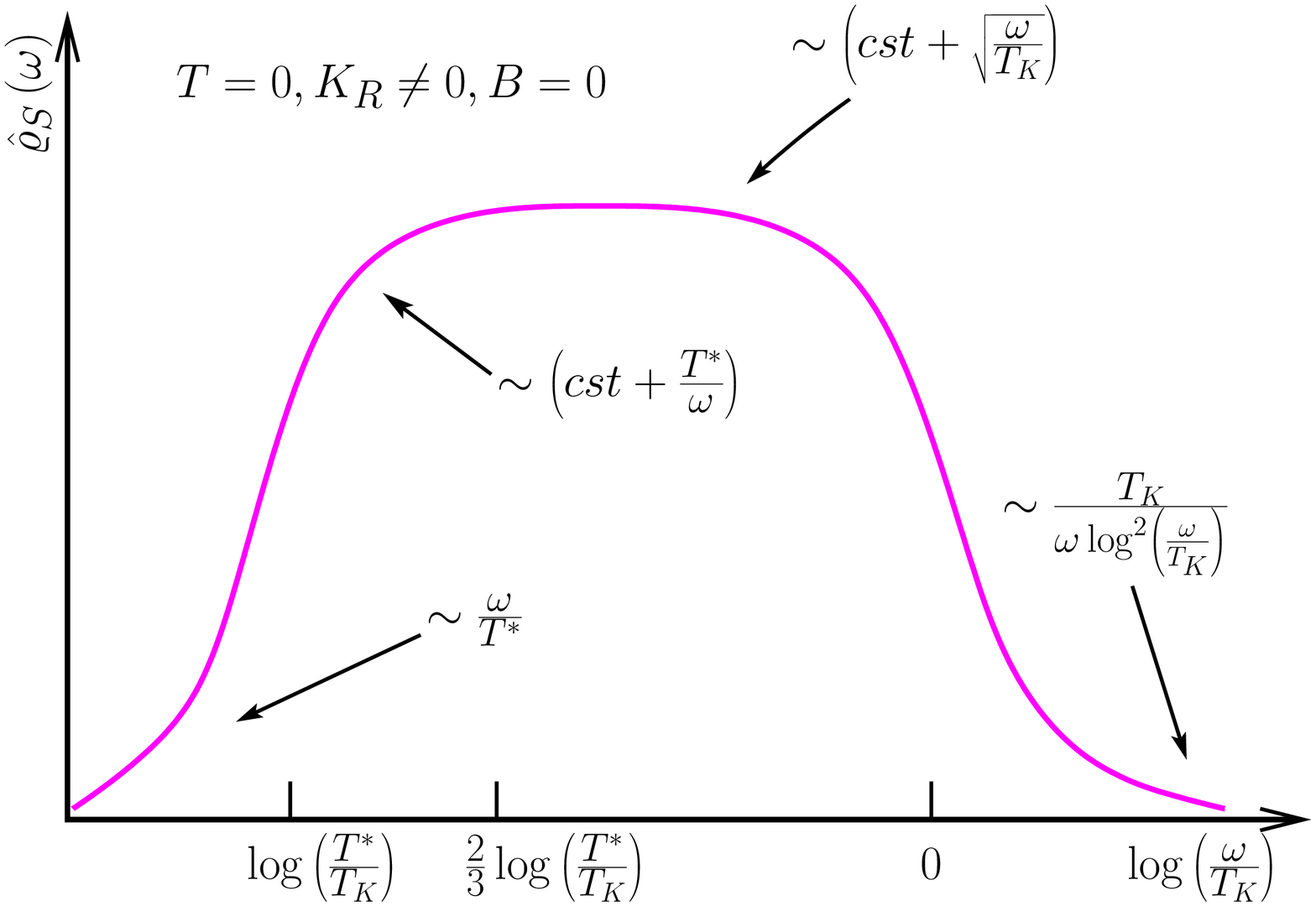}
  \caption{(color online) Top: 
    Sketch of the dimensionless spectral function of $\vec S$:
    $\hat{\varrho}_S=T_K{\varrho}_S=-T_K\im \chi_S(\omega)/\pi$ for $T>0$ and $K_R=0,B=0$ as a function of
    $\log\left(\omega/T_K\right)$.
    Bottom: Sketch of $\hat{\varrho}_S=T_K{\varrho}_S=-T_K\im \chi_S(\omega)/\pi$ for $T=0$ and $K_R\ne0,B=0$ as a function of
    $\log\left(\omega/T_K\right)$.
    Asymptotics indicated for $\omega<T_K$ were derived through scaling
    arguments. The large $\omega$-behavior is a result of perturbation theory.\cite{Garst_05}
  }
  \label{fig:rho_Simp_finT_finTast_sketch}
\end{figure}

We shall not repeat here all the steps of the derivation, only summarize  the
main results. In the absence of a magnetic field, $h=0$ the spectral function
of the spin operator is odd. Furthermore, at $T=0$ and for no anisotropy, $\kappa
= 0$, 
the spectral function has a jump at $\omega=0$,\cite{Cox_97}
\be
\hat \varrho_S^{T,h,\kappa=0}(\omega) \approx {\rm sgn}(\omega) \Bigl[r_S
+ r_S^{\;\prime}
\; 
\sqrt{\frac{|\omega|}{T_K}}
+\dots\Bigr]\;.
\ee 
This jump corresponds  to a logarithmically divergent dynamical susceptibility, 
$\re \chi_S(\omega) = -\re {\cal G}_S(\omega)\propto \ln(T_K/\omega)/T_K  $.

For $\omega\gg T_K$ the impurity spin becomes asymptotically free, decoupled
from the conduction electrons, therefore its $\omega$-dependence is set  by
its scaling dimension at the free fermion fixed point where $x_S^{free}=0$. It has the implication
that its correlation fuction decays as $\omega^{-1}$ corresponding to the
Curie-Weiss susceptibility with logarithmic corrections present, known from
Bethe Ansatz results and from perturbation theory. 

At finite temperatures  $T\ne 0$, but for $\kappa=h=0$, we obtain the following
scaling form for $T,\omega\ll T_K$:  
\begin{multline}
\hat \varrho_S
^{h,\kappa=0}(\omega)
\equiv\Theta_{ S}\left(\frac{\omega}{T}\right)+
\sqrt{\frac{T}{T_K}} \;\;
{\tilde
  \Theta}_{ S}\left(\frac{\omega}{T}\right)+\dots.
\end{multline}
The asymptotic properties of the scaling functions 
$\Theta_{ S}$ and ${\tilde\Theta}_{ S}$  are listed in
Table~\ref{tab:asymp}. 

\begin{figure}[t]
  \includegraphics[width=0.9\columnwidth,clip]{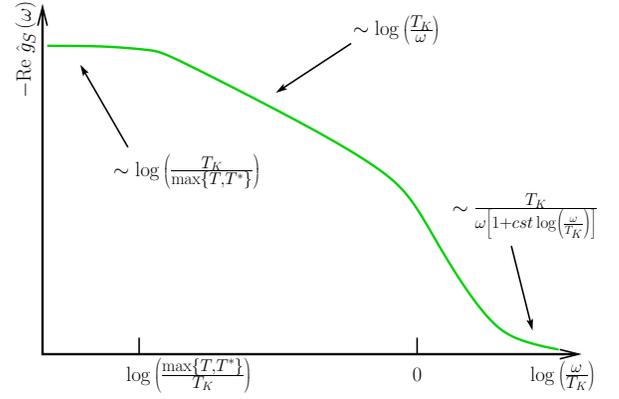}
  \caption{(color online) Sketch of the real part of the dimensionless
    Green's function of ${\vec S}$,  
    $\re {\hat g}_S= T_K\;
    \re\chi_S(\omega)\equiv T_K\;\re\G_S(\omega)\;$ for $T,T^*> 0$ 
    as a function of $\log\left(\omega/T_K\right)$. 
  }
  \label{fig:re_g_S_fin_T_Tstar}
\end{figure}
In case of finite channel anisotropy but zero temperature we obtain
for $\omega\ll T_K$ the scaling form
\begin{multline}
\hat\varrho_{S}^{T,h=0}(\omega) \approx \K_{
  S}\left(\frac{\omega}{T^*}\right)+
\sqrt{\frac{|\omega|}{T_K}}\;\;
{\tilde\K}_{
  S}\left(\frac{\omega}{T^*}\right)+\dots\;.
\end{multline}
The asymptotic properties of $\K_{ S},{\tilde\K}_{ S}$ are only slightly different
from those of $\Theta_{ S},{\tilde\Theta}_{ S}$ (see Table~\ref{tab:asymp}):
below $T^\ast$ the spectral function displays analytic behavior, while 
the regime $\omega> T^*$ is governed by non-analytical corrections associated
with the 2CK fixed point. 
 In this regime a feature 
worth mentioning is the appearance of a correction,  $\sim T^*/\omega$ 
to  $\K_{ S}$, more precisely, the lack of a 
$\sqrt{|T^*/\omega|}$ correction. 
This is due to the fact that the anisotropy operator is odd, while the spin
operator is even with respect to swapping the channel labels. 
Therefore there is no first order correction to  the spin-spin correlation 
function in $\kappa$, and the leading corrections are only of second order,
i.e., of the form  $\kappa^2/\omega$. From the comparison of the  
terms in
$\K_{
  S}$ and  ${\tilde\K}_{ S}$ 
it also follows the existence of another 
cross-over scale, 
\be 
 T_s^{**}\sim\left({T^\ast}^2T_K\right)^{1/3},
\ee 
that separates the  regimes governed by the leading relevant and leading irrelevant operators. 
Here we used the subscript $s$ to indicate that
this scale $T_s^{**}$ is different from the scale $T_f^{**}$ introduced in
relation to the  local fermion's spectral function. 
The asymptotic properties of ${\hat\varrho}_{ S}\propto\chi_S(\omega)$ for 
$T>0,K_R=0$ and $T=0,K_R\neq0$ are sketched in the upper and lower parts of 
Fig.~\ref{fig:rho_Simp_finT_finTast_sketch}, while the 
behavior of the real part is 
presented in Fig.~\ref{fig:re_g_S_fin_T_Tstar}.

\begin{figure}[ht]
  \includegraphics[width=0.9\columnwidth,clip]{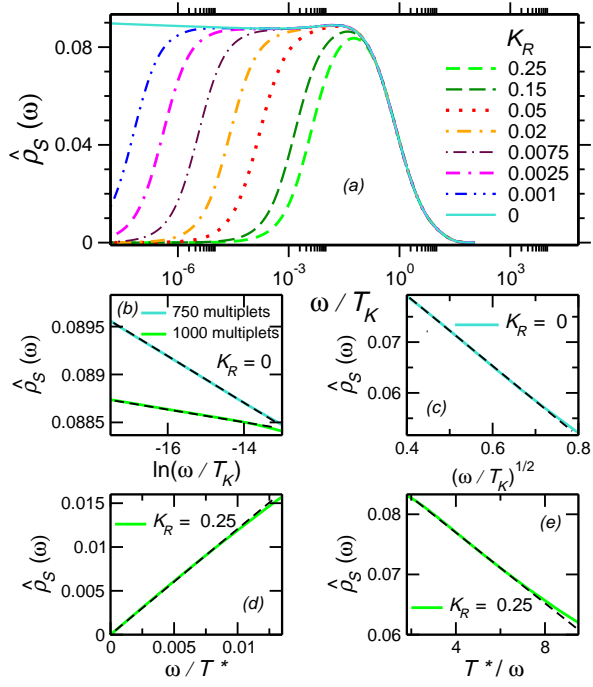}
  \caption{(color online) $(a)$ Dimensionless spectral function of 
    ${\vec S}$: ${\hat\varrho}_S=T_K{\varrho}_S=-T_K\im\chi_S(\omega)/\pi$ as a function of $\omega/T_K$ for
    different values of $K_R$. $(b)$ Minute $\log\left(\omega\right)$-dependence
    at the lowest frequencies diminishing as a function of the number of kept multiplets.
    $(c-e)$ Numerical confirmations of the low-frequency asymptotics derived
    from scaling arguments in Section \ref{sec:spin}. 
Straight dashed lines are to demonstrate deviations from the expected $\sqrt{\omega}$-like $(c)$, 
    ${\omega}$-like $(d)$, and $1/\omega$-like 
behavior $(e)$. In plots $(d-e)$ $T^\ast/T_K=7\times 10^{-2}$.
  }
  \label{fig:rho_S_chani_TK_cft}
\end{figure}
The expectations above are indeed nicely born out  by the NRG calculations:
Fig.~\ref{fig:rho_S_chani_TK_cft} shows the impurity spin spectral
functions as a function of $\omega/T_K$  for various $K_R$-s and  
their asymptotic properties. 
First, in Fig.~\ref{fig:rho_S_chani_TK_cft}.$(b)$  
we show a very small  logarithmic $\omega$-dependence that 
we observed  below $T_K$ at the 2CK fixed point.  The amplitude of this
$\log(\omega)$-dependence  
 was reduced as  we increased 
the number of multiplets. 
It appears that this behavior is not derived from the lognormal
smoothing of the NRG data, and it may be due to some approximations used in
the spectral sum-conserving DM-NRG 
procedure. In Fig.\ \ref{fig:rho_S_chani_TK_cft}.$(c)$
we show the square root-like behavior around the 2CK Kondo fixed point which
is attributed to the leading irrelevant operator, while 
Fig.\ \ref{fig:rho_S_chani_TK_cft}.$(d)$ shows that first order
corrections  coming from the scaling of the channel 
anisotropy are indeed absent just as we stated above, and  only second order terms appear, resulting in an
$1/\omega$-like behavior. Finally,  Fig.\ \ref{fig:rho_S_chani_TK_cft}.$(e)$
demonstrates the linear $\omega$-dependence, 
 which is characteristic of most bosonic operators in
the proximity of an FL fixed point. All these findings support very nicely the 
analytical properties summarized in  Table~\ref{tab:asymp}.

\begin{figure}[t]
  \includegraphics[width=0.9\columnwidth,clip]{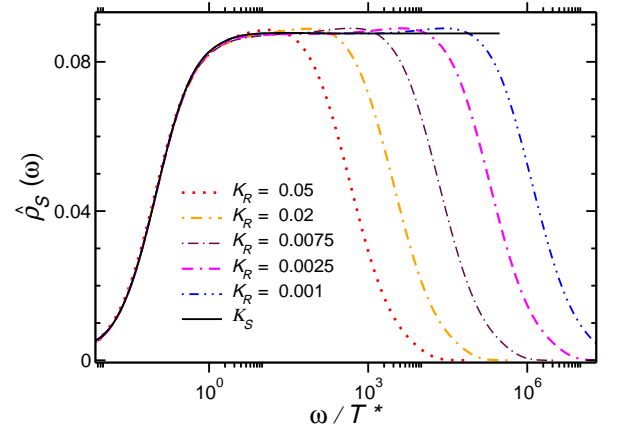}
  \caption{(color online) Universal collapse of the dimensionless spectral 
    function of ${\vec S}$:  ${\hat\varrho}_S=T_K{\varrho}_S$ to the scaling curve, 
    $\K_S$ as a function of $\omega/T^\ast$ for sufficiently small, 
    non-zero values of $K_R$.
  }
  \label{fig:rho_S_chani_scaled}
\end{figure}
The spin spectral functions also collapse to a universal scaling curve describing the cross-over from the 
two-channel Kondo to the single channel Kondo fixed points, when they are
plotted against $\omega/T^{**}$. 
This  universal data collapse 
is  demonstrated in Fig.~\ref{fig:rho_S_chani_scaled} where the impurity
spin spectral functions are plotted 
  for various $K_R$ values. The data collapse works up to 
somewhat higher anisotropy values  than 
for  the local fermions' spectral functions as it is indicated by 
the $K_R$-dependence of the scales $T^{**}_s$ and $T^{**}_f$.

\begin{figure}[b]
  \includegraphics[width=0.9\columnwidth,clip]{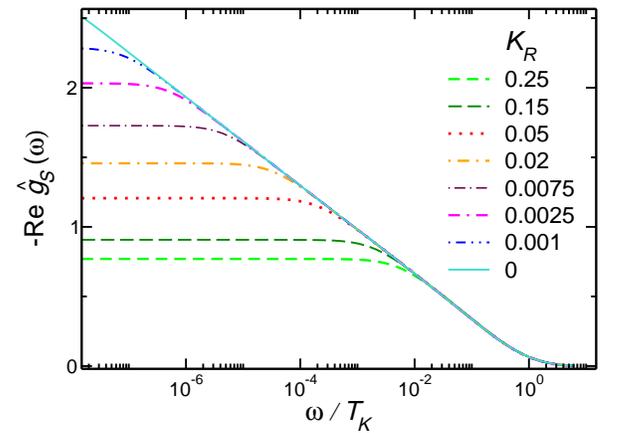}
  \caption{(color online) Real part of the dimensionless Green's function 
    (susceptibility) of 
    ${\vec S}$, $\re {\hat g}_S = - T_K\;
    \re\chi_S(\omega) $, 
    as a function of $\omega/T_K$, for different values of $K_R$.
  }
  \label{fig:re_g_S_chani_TK}
\end{figure}

The real part of the spin susceptibility was obtained through 
numerical Hilbert transformation, and is shown in Fig.~\ref{fig:re_g_S_chani_TK} 
as a function of $\omega/T_K$ for various values of $K_R$. These curves meet 
the expected behavior sketched in Fig.~\ref{fig:re_g_S_fin_T_Tstar}: they
display a logarithmic increase at high-frequencies and saturate at values 
that correspond to $\re\chi_S\sim {\rm ln}(T_K/T^*)\; /\; T_K$.

Let us finally discuss the  case, $T= \kappa=0$ but $h\neq 0$. Then the
components of  ${\vec S}$ are distinguished by the magnetic field: The
spectral function of $S^z$ has almost the same features as for finite
channel anisotropies. Since $S^z$ is a hermitian  
operator, its spectral function remains odd  and acquires the following
corrections in the different scaling regimes
\begin{multline}
\hat \varrho_{S,z}
\equiv\B_{S,z}\left(\frac{\omega}{T_h}\right)+
\sqrt{\frac{|\omega|}{T_K}}\;\;
{\tilde\B}_{S,z}\left(\frac{\omega}{T_h}\right)+\dots
\end{multline}
with the scaling functions $\B_{S,z},{\tilde\B}_{S,z}$ having the 
asymptotic properties listed in Table~\ref{tab:asymp}. 

Note that in this case the first order correction coming from  the magnetic
field does not vanish, and  leads to  the appearance of a 
cross-over scale $\sim \sqrt{T_hT_K}$.

The perpendicular components of the impurity spin have somewhat different properties.  
First of all, the operators $S^\pm$ are not Hermitian, and therefore their
spectral functions are not symmetrical. The spectral functions of the 
operators $S^x$ and $S^y$ are, however, symmetrical, and their  
Green's functions (and susceptibilities) are related through  
\bea 
{\cal G}_S^x = 
{\cal G}_S^y = 
\frac 1 4 ( {\cal G}_S^{+-} +  {\cal G}_S^{-+})\;.
\eea

\begin{figure}[t]
  \includegraphics[width=0.9\columnwidth,clip]{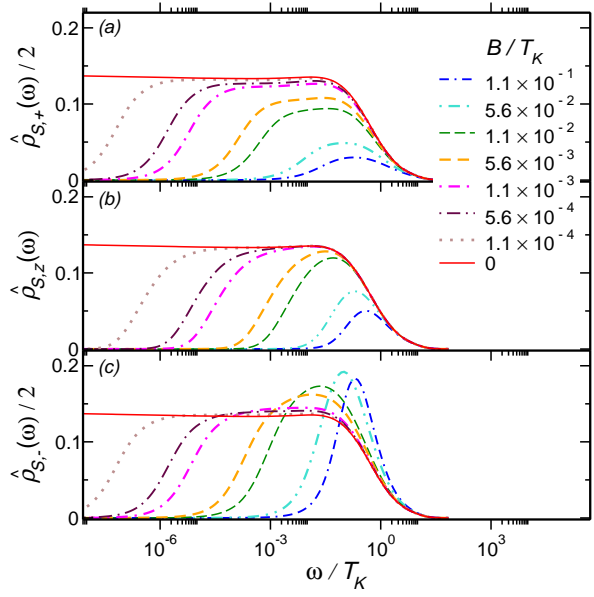}
  \caption{(color online) $(a)$ Dimensionless spectral function of 
    ${S}^{+}$: ${\hat\varrho}_{S,+}=T_K{\varrho}_{S,+}$, $(b)$ of ${S}^{z}$:
    ${\hat\varrho}_{S,z}=T_K{\varrho}_{S,z}$ and $(c)$ of ${S}^{-}$:
    ${\hat\varrho}_{S,-}=T_K{\varrho}_{S,-}$ for different 
    values of $B$ as a function of $\omega/T_K$.
  }
  \label{fig:rho_S_Bneq0}
\end{figure}
\begin{figure}[t]
  \includegraphics[width=0.9\columnwidth,clip]{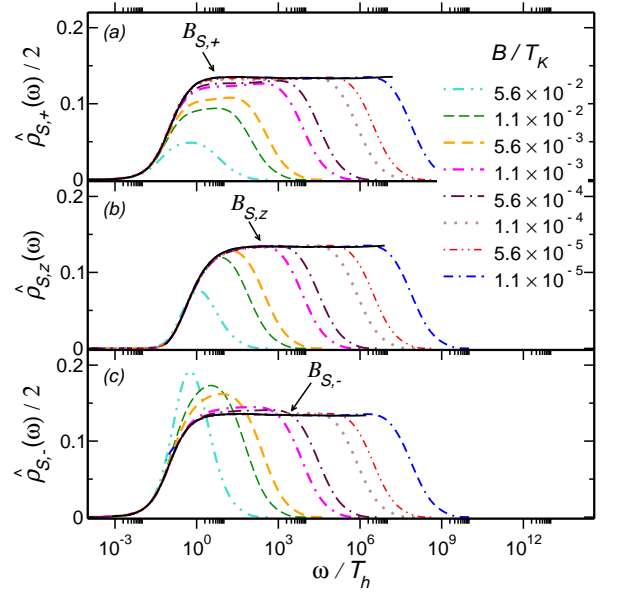}
  \caption{(color online) Universal collapse of 
    ${\hat\varrho}_{S,+}=T_K{\varrho}_{S,+}$, ${\hat\varrho}_{S,z}=T_K{\varrho}_{S,z}$ and 
    ${\hat\varrho}_{S,-}=T_K{\varrho}_{S,-}$ to the three scaling curves: $\B_{S,+},\B_{S,z}$
    and $\B_{S,-}$ for sufficiently small, non-zero values of $B$ as a 
    function of $\omega/T_h$.
  }
  \label{fig:rho_S_Bneq0_scaled}
\end{figure}

The corresponding dimensionless 
spectral functions,  $\hat \varrho_s^z$ and  $\hat \varrho_s^\pm$
as computed by our DM-NRG calculations are shown in Fig.\
\ref{fig:rho_S_Bneq0} 
as a function of $\omega/T_K$, while  the universal scaling  with $\omega/T_h$
is confirmed for low-frequencies in Fig.~\ref{fig:rho_S_Bneq0_scaled}.  
This scaling also turned out to be valid for values of $B$ higher
than the ones for fermions (see Fig.\
\ref{fig:rho_S_Bneq0_scaled}).
The scaling functions ${\cal B}_{S,z}$ and
${\cal B}_{S,\pm}$ behave very similarly. This is somewhat surprising, since
the naive expectation would be to have a {\em resonance} in ${\cal B}_{S,+}$, 
just as in the local fermion's spectral function, that
would correspond to a spin-flip excitation  at the
renormalized spin splitting, $T_h$. 

excitations''.   

\begin{figure}[t]
  \includegraphics[width=0.9\columnwidth,clip]{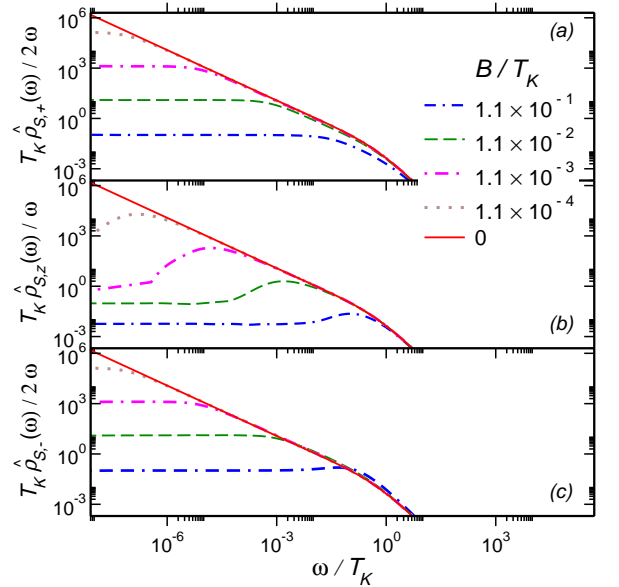}
  \caption{(color online) $(a)$ $\varrho(\omega)/\omega$ of 
    ${S}^{+}$: ${T_K\hat\varrho}_{S,+}/2\omega=T^2_K{\varrho}_{S,+}/2\omega$, $(b)$ of ${S}^{z}$:
    ${T_K\hat\varrho}_{S,z}/\omega=T^2_K{\varrho}_{S,z}/\omega$ and $(c)$ of ${S}^{-}$:
    ${T_K\hat\varrho}_{S,-}/2\omega=T^2_K{\varrho}_{S,-}/2\omega$ for different 
    values of $B$ as a function of $\omega/T_K$.
  }
  \label{fig:rho_S_per_omega}
\end{figure}

However, quite remarkably, a resonance seems to appear in
$\chi_{S,z}^"(\omega)/\omega$
at a frequency $\omega\sim T_h$,   while we find no resonance  
in $\chi_{S,\pm}^"(\omega)/\omega$. This can be seen in  
Fig.\ \ref{fig:rho_S_per_omega}, where  $T_K^2\varrho(\omega)/\omega$ is plotted
for the different spin components as a function of $\omega/T_K$ for various
magnetic field values. This seems to indicate that the spin coherently oscillates between
the spin up and spin down components, while its $x,y$ components simply relax
to their equilibrium value.

\section{Superconducting correlations}\label{sec:superC}
In the last section, let us investigate the 
local superconducting  correlation functions. These deserve special 
attention, since many heavy fermion compounds display exotic superconducting
phases that may possibly be 
induced  by local two-channel Kondo physics.\cite{Cox_97} 
The 
 most obvious candidates for the corresponding local operators have been
identified in Section~\ref{sec:NFL_fixed_point}, and are the 
local channel-asymmetric superconducting 
operator, ${\cal O}_{SC} = \; f^\dagger_{0,1,\uparrow}f^\dagger_{0,2,\downarrow} -
  f^\dagger_{0,1,\downarrow} f^\dagger_{0,2,\uparrow}$, and 
the composite fermion superconductor field, 
${\cal O}_{SCC} = f^\dagger_{0,1}
\vec S \vec \sigma\; i\sigma_y f^\dagger_{0,2}$. 

For the composite superconductor we find the expansion, 
\be
{\cal O}_{SCC}  =  A_{SCC}\;  \phi_\Delta^{++} + \dots\;
\ee
where the expansion coefficient  $A_{SCC}$ can be estimated from the 
high-frequency behavior of the correlation function up to logarithmic prefactors
as  $A_{SCC}\sim \sqrt{T_K}/D_F$. While for the impurity spin, one can 
exclude logarithmic corrections to the expansion coefficient $A_S$ in Eq.\ \ref{eq:spin_coeff}
 based upon the exact Bethe Ansatz results, 
this is not possible for the superconducting correlation function.
In fact, we know that in the expansion of the composite
fermion itself the correct prefactor is $A_F\sim J/\sqrt{T_K}\sim
1/\bigl(\sqrt{T_K}\ln(D_F/T_K)\bigr)$.\cite{Costi_00} Therefore,  
similar logarithmic factors could appear in the prefactor
$A_{SCC}$. Nevertheless, in the following, we shall disregard possible
logarithmic corrections, and define the normalized dimensionless 
and scale-invariant correlation function through the relation, 
\be 
 \frac {D_F^2}{T_K} \;{\cal G}_{SCC}(\omega) =  {\hat g}_{SCC}(\omega)\;.
\ee
Apart from 
its overall amplitude and 
 its high-frequency behavior, 
in the low-frequency scaling regimes the spectral
function  of the composite
superconductor operator behaves the same
way as that of ${S}^z$ (see Tab.\ \ref{tab:conf_weights}). 
Therefore we merely state its
asymptotics without further explanation.

In the absence of anisotropy and magnetic field, $\kappa=h=0$, for $\omega \ll
T_K$ the spectral function becomes a universal function, $\hat
\rho_{SCC}(\omega/T)$, whose behavior  is described by the scaling form, 
\begin{multline}
 \hat \varrho^{h,\kappa=0}_{SCC}(\omega)
\approx \Theta_{
  SCC}\left(\frac{\omega}{T}\right)+
\sqrt{\frac{T}{T_K}}
\;{\tilde
\Theta}_{SCC}\left(\frac{\omega}{T}\right)+\dots\;,
\end{multline}
while 
in the presence of anisotropy, but at $T=0$ temperature and for $h=0$,
the spectral functions behave as 
\begin{multline}
\hat\varrho^{T,h=0}_{SCC}(\omega)  \approx \K_{
  SCC}\left(\frac{\omega}{T^*}\right)+
\sqrt{\frac{|\omega|}{T_K}}\;  
{\tilde\K}_{
  SCC}\left(\frac{\omega}{T^*}\right)+\dots\;.
\end{multline}

\begin{figure}[t]
  \includegraphics[width=0.9\columnwidth,clip]{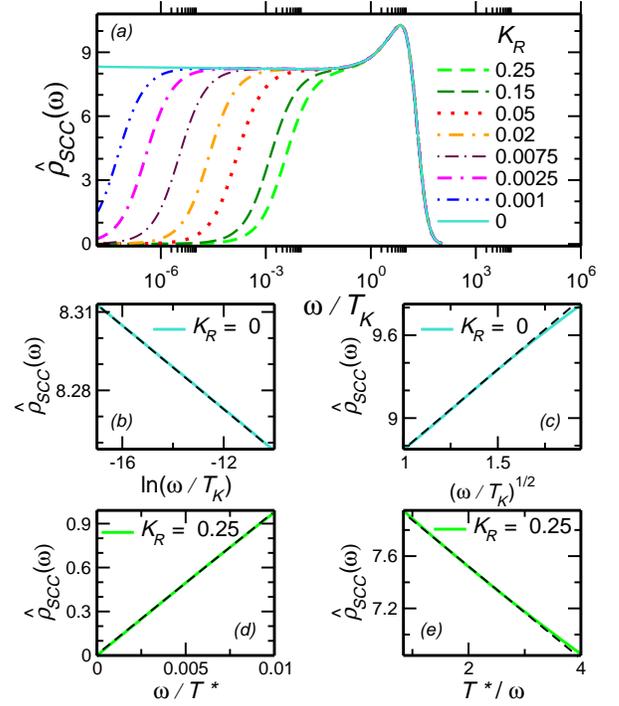}
  \caption{(color online) $(a)$ Dimensionless spectral function,
    ${\hat\varrho}_{SCC}=D_F^2/T_K\;{\varrho}_{SCC}$ of the operator  
    ${\cal O}_{SCC}$, as a function of $\omega/T_K$ for
    different values of $K_R$. $(b)$. The very weak 
    $\log\left(\omega\right)$-dependence
    at the lowest frequencies. This dependence
    is suppressed  as 
    we increased 
    the number of kept
    multiplets.
    $(c-e)$ Numerical confirmations of the low-frequency asymptotics derived
    from scaling arguments in Section \ref{sec:superC}. 
    Dashed straight lines are to demonstrate deviations from the expected $\sqrt{\omega}$-like $(c)$, 
    ${\omega}$-like $(d)$ and $1/\omega$-like $(e)$ behavior. In plots $(d-e)$
    $T^*/T_K=7\times 10^{-2}$.
  }
  \label{fig:rho_SCC_chani_TK_cft}
\end{figure}
Finally, in a finite magnetic field but for $\kappa=0$ anisotropy and  
$T=0$ temperature the spectral function assumes the following scaling form, 
\be
\hat \varrho_{SCC}^{\kappa=T=0}
\equiv\B_{SCC}\left(\frac{\omega}{T_h}\right)
+ 
\sqrt{\frac{|\omega|}{T_K}}
\;\;{\tilde\B}_{SCC}\left(\frac{\omega}{T_h}\right)+\dots\;.
\ee 
The properties of the the various scaling functions defined above are
identical to those of the corresponding spectral functions of 
the $S^z$, which were detailed in Table~\ref{tab:asymp}, therefore they have
not been 
included in Table~\ref{tab:asymp}.

\begin{figure}[t]
  \includegraphics[width=0.9\columnwidth,clip]{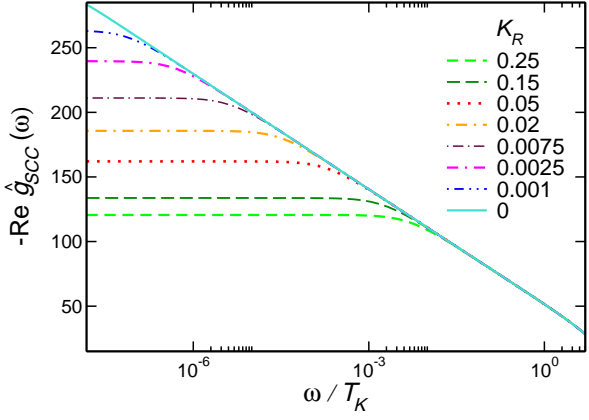}
  \caption{(color online) Real part of the dimensionless Green's function 
    of ${\cal O}_{SCC}$: $\re {\hat g}_{SCC}=D_F^2/T_K\;\re\G_{SCC}$ as a function of $\omega/T_K$ 
    for different values of $K_R$.
  }
  \label{fig:re_g_SCC_chani_TK}
\end{figure}

\begin{figure}[b]
  \includegraphics[width=0.8\columnwidth,clip]{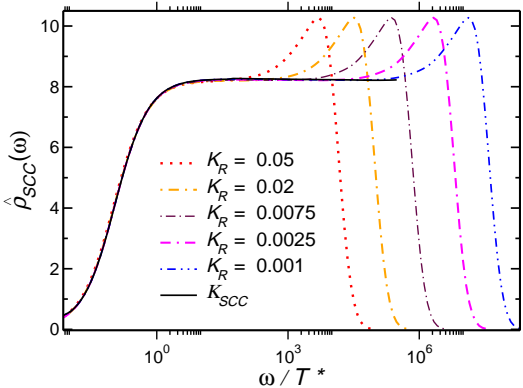}
  \caption{(color online) Universal collapse of ${\hat\varrho}_{SCC}$ 
    to the scaling curve $\K_{SCC}$ as a function of
    $\omega/T^\ast$ for sufficiently small, non-zero values of $K_R$.
  }
  \label{fig:rho_SCC_chani_scaled}
\end{figure}

The asymptotic properties are nicely confirmed by our NRG calculations.
The dependence on the anisotropy, together with the  
$\sim \sqrt{|\omega|}$, the $\sim 1/\omega$ and the $\sim \omega$ 
scaling regimes are plotted in Fig.~\ref{fig:rho_SCC_chani_TK_cft}. 
Here the high-frequency region, $\omega>T_K$, is also displayed, 
where the spectral function is roughly linear in the frequency, 
as dictated by the free fermion fixed point. 

Fig.~\ref{fig:re_g_SCC_chani_TK} displays the real part of the 
dimensionless Green's function, that is essentially the real part of the
superconducting susceptibility. This diverges logarithmically
for $T^*=0$, but for finite $T^*$'s it saturates, corresponding to a
susceptibility  value
$$
\chi_{SCC}\sim \frac{T_K}{D_F^2} \ln\left(\frac{T_K}{T^*}\right)\;.
$$
Notice that there is a small prefactor in front of the logarithm that
arises from the asymptotically free behavior at large frequencies.

\begin{figure}[t]
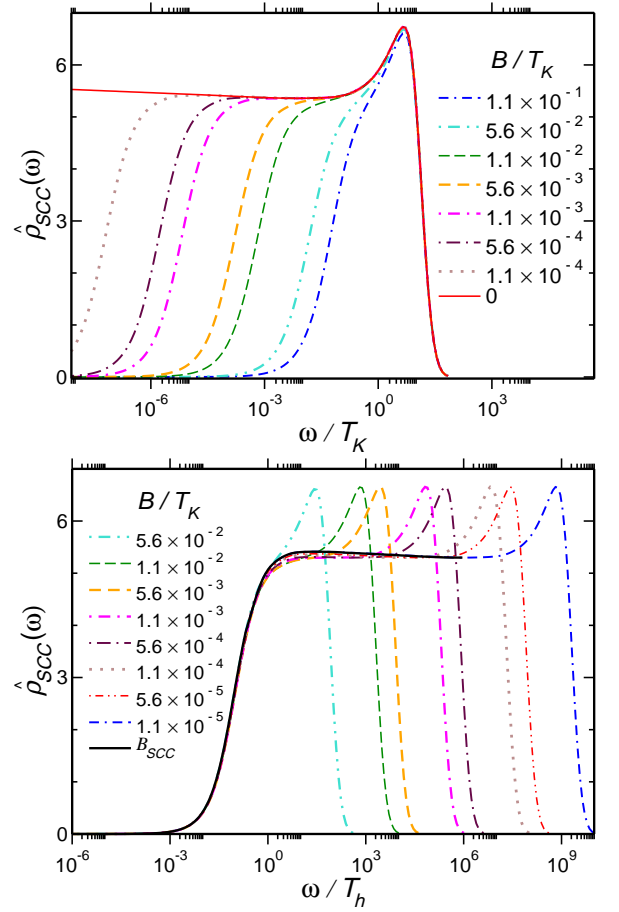

  \includegraphics[width=0.9\columnwidth,clip]{fig_6_RR/rho_SCC_Bneq0_RR}
  \includegraphics[width=0.9\columnwidth,clip]{fig_6_RR/rho_SCC_Bneq0_scaled_RR}
  \caption{(color online) Top: Dimensionless spectral function 
${\hat\varrho}_{SCC} = -(D_F^2/\pi T_K)\;\im\chi_{SCC}= -(D_F^2/\pi T_K)\;\im\G_{SCC}$ 
of the composite superconductor operator  
${\cal O}_{SCC}$  for different values of $B$, as a function of $\omega/T_K$.
Bottom: Universal collapse of ${\hat\varrho}_{SCC}$ to the scaling curve $\B_{SCC}$ 
    for sufficiently small, non-zero values of $B$ as a function of $\omega/T_h$.
  }
  \label{fig:rho_SCC_Bneq0}
\end{figure}
The universal collapse of the low-frequency part of the curves 
in terms of $\omega/T^*$ is shown in Fig.~\ref{fig:rho_SCC_chani_scaled}. 
The cross-over curve, $\K_{SCC}\left(\frac{\omega}{T^*}\right)$ is very similar
to the  spin cross-over function, $\K_{S}$, and displays a
plateau at large frequencies from which it deviates as $1/\omega$, 
until it finally reaches the linear frequency regime below $T^*$. 

Application of a magnetic field has effects very similar to the anisotropy, as
shown in the upper part of Fig.~\ref{fig:rho_SCC_Bneq0}. In Fig.~\ref{fig:rho_SCC_Bneq0} the
small logarithmic increase at small frequencies is more visible. As mentioned
before, this increase
is most likely  an artifact of  the spectral sum conserving approximation of
Ref.~\onlinecite{DMNRG}
and it is due to the way this method 
redistributes spectral weights. This is based on the observation that the 
 slope of the logarithm gets smaller if we increase the  number of 
 multiplets kept. These curves also collapse to a single universal curve
as a function of $\omega/T_h$, as shown in the lower part of 
Fig.~\ref{fig:rho_SCC_Bneq0}.

\begin{figure}[t]
  \includegraphics[width=1.\columnwidth,clip]{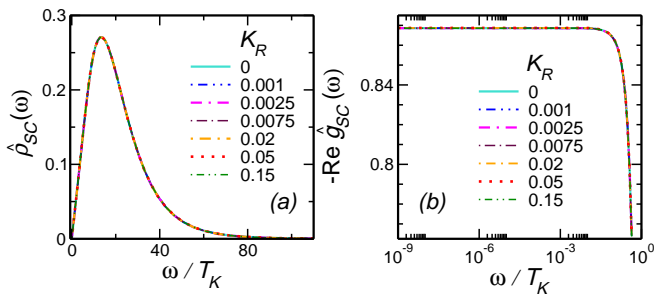}
  \caption{(color online) $(a)$ Dimensionless spectral function of ${\cal O}_{SC}$: ${\hat\varrho}_{SC}=D_F\;{\varrho}_{SC}$ as a function of
    $\omega/T_K$ for different  values of $K_R$, and
    $(b)$ the real part of its dimensionless Green's function: $\re\hat{g}_{SC}=D_F\;\re\G_{SC}$.
  }
  \label{fig:SC_chani_rho_re_g_TK}
\end{figure}

Finally, in Fig.~\ref{fig:SC_chani_rho_re_g_TK},   we show the numerically
obtained spectral function and the corresponding dimensionless susceptibility
of the non-composite superconductor,  ${\cal O}_{SC}= 
f^\dagger_{0,1,\uparrow}f^\dagger_{0,2,\downarrow} -
  f^\dagger_{0,1,\downarrow} f^\dagger_{0,2,\uparrow}$. Clearly, this
spectral function displays no plateau below $T_K$, but it exhibits a linear in
$\omega$ behavior below $T_K$, and correspondingly, the 
susceptibility 
$\re\chi_{SC}$ remains finite for $\omega\to 0$ even in the absence of
anisotropy and an external magnetic field, i.e., at the 2CK fixed point. 

This implies that, although its charge and spin quantum numbers would allow
it, the expansion of this operator does {\rm not} contain the scaling operator
$\phi^{\tau\tau'}_\Delta$. This may be due to the difference 
 in the Ising
quantum numbers, which we 
 did not  identify. Thus the 
dimension of the highest-weight scaling operator that appears in the expansion 
of  ${\cal O}_{SC}$ is $x=1$ and not 1/2, as one would naively expect
based upon a simple comparison of quantum numbers.
Turning on a small anisotropy or  magnetic field does not influence
substantially the spectral properties of the corresponding Green's function, either. 

\section{Conclusions}\label{sec:fin}

In the present paper we gave a detailed discussion of  the spectral properties of the 
two-channel Kondo model. We analyzed the properties of the correlation
functions of various local operators in the presence of a channel anisotropy
and an external magnetic field. In particular, we studied numerically and
analytically the correlation functions of the local fermions, 
 $f_{\alpha,\sigma}\equiv
f_{0,\alpha,\sigma}$, the components of the impurity spin, $\vec S$, 
the local superconductivity operator, ${\cal O}_{SC} \equiv
f^\dagger_{1}i\sigma_y f^\dagger_{2}$,  and the 
composite superconductor operator, $f^\dagger_{1} \vec S \vec \sigma\;
i\sigma_yf^\dagger_{2}$. 
The selection of these operators was partially motivated by 
conformal field theory, which tells us  the quantum 
numbers and scaling dimensions of the various scaling 
operators at the two-channel Kondo fixed point.\cite{Affleck_Ludwig_91}
There are, however, many operators that have quantum numbers identical with the
scaling fields.  Here we picked operators having the right quantum numbers, 
 and at the same time
having the largest possible scaling dimension at the free fermion fixed point,
 where
$J_1,J_2\to0$. These are the operators, whose spectral functions  are expected to have the largest
spectral weight at small temperatures (among those having the same quantum
numbers), and which are therefore the primary candidates for 
an order parameter, when a lattice of 2CK impurities is formed, as is the case
in some Uranium and Cerium-based compounds.  The operators above are, of course, 
also of physical interest on their own: the spectral function of $f_{\alpha,\sigma}$ is related
to the tunneling spectrum 
into the conduction electron see at the impurity site, 
the Green's function of $\vec S$ is just the dynamical spin susceptibility
that can be measured under inelastic neutron scattering, and finally
the local superconducting operators are candidates for superconducting ordering 
in heavy fermion materials. We remark that, in the electron-hole symmetrical
case,  the other components of the operator multiplet that contains the
composite superconducting order parameter  $O_{SCC}$ would correspond to a
composite  channel-mixing  charge density ordering.
Of course, the susceptibilities of this operator has the same properties as
 that of 
$\chi_{SCC}(\omega)$. 
 
In addition to these operators, there are two more operators of possible
interest:  the so-called  composite Fermion's Green's function is related to
the $T$-matrix, $T(\omega)$ that describes the
scattering properties off a two-channel impurity (or the conductance through
it in case of a quantum dot), and  was already  studied in detail 
in Ref.~\cite{2ck_cond}. A further candidate is the channel anisotropy 
operator. This has also a logarithmically  divergent susceptibility, 
and would also be associated with a composite orbital ordering  in case of 
a two-channel Kondo lattice system. However, the spectral properties of this
latter 
 operator are so similar to those of the composite superconductor
that we have decided no to show data about them.

For the numerical calculations we used a flexible DM-NRG method, where 
 we exploited  the hidden charge SU(2) symmetries\cite{Jones_87} 
as well as 
the 
invariance  under spin rotations to obtain
high precision data. To identify the scaling operators
in this case, we reconstructed the boundary conformal field theory of Affleck and
Ludwig for this symmetry classification. We then established the 
scaling properties of the various dynamical correlation functions 
and identified the corresponding universal cross-over functions and their
asymptotic properties, based upon simple but robust
scaling arguments. In this way, universal scaling functions describing the 
cross-over from the two-channel Kondo fixed point to the single channel Kondo
fixed point (for $J_1\ne J_2$) and to the magnetically polarized fixed point
(for $B\ne 0$) have been introduced, which we then determined numerically. 
We emphasize again that presently these universal cross-over functions can only be determined 
through the application of DM-NRG, and in fact, for the scaling curves in the
presence of a magnetic field the application of the DM-NRG method was
absolutely necessary. 

\begin{figure}[t]
  \includegraphics[width=0.75\columnwidth,clip]{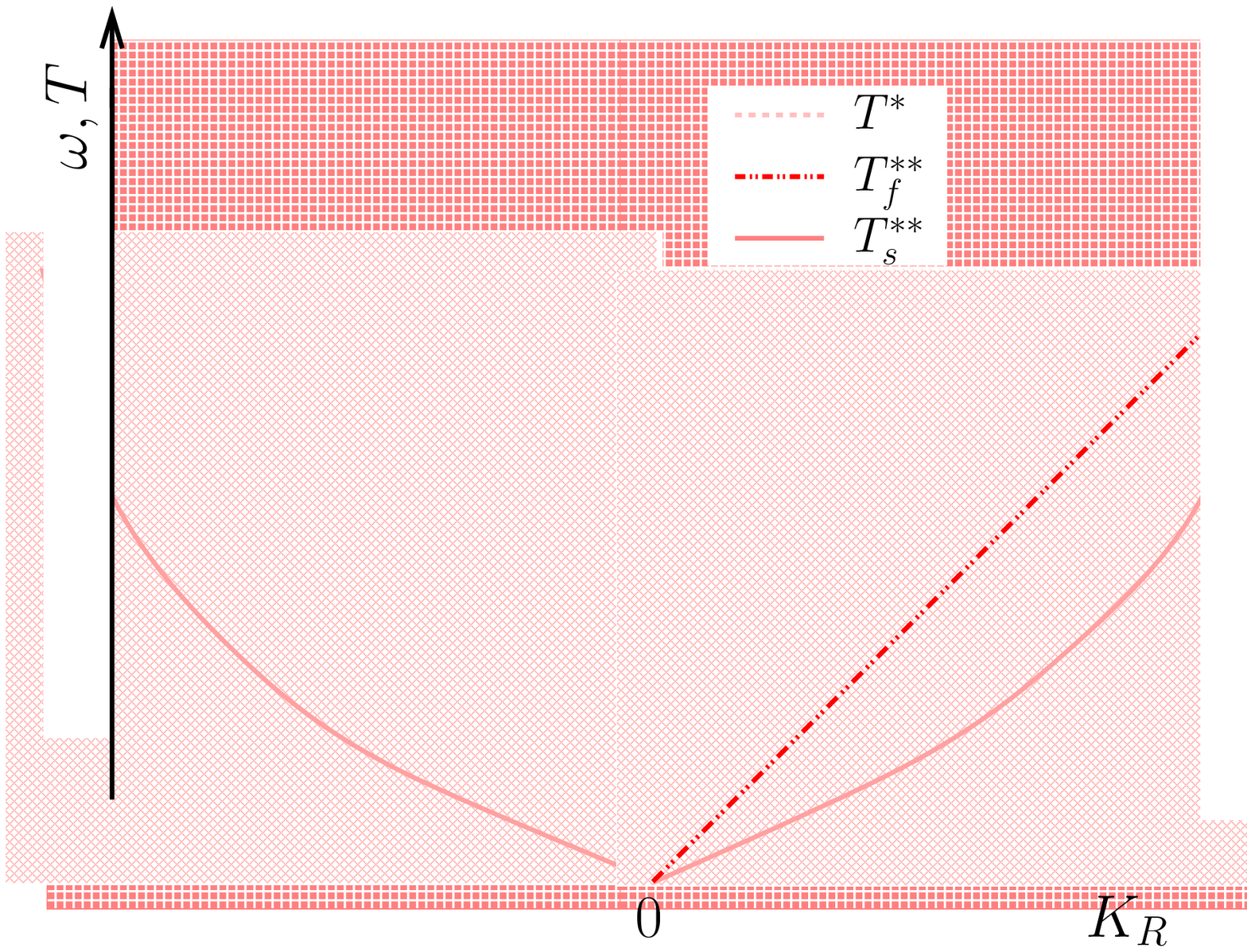}
  \includegraphics[width=0.75\columnwidth,clip]{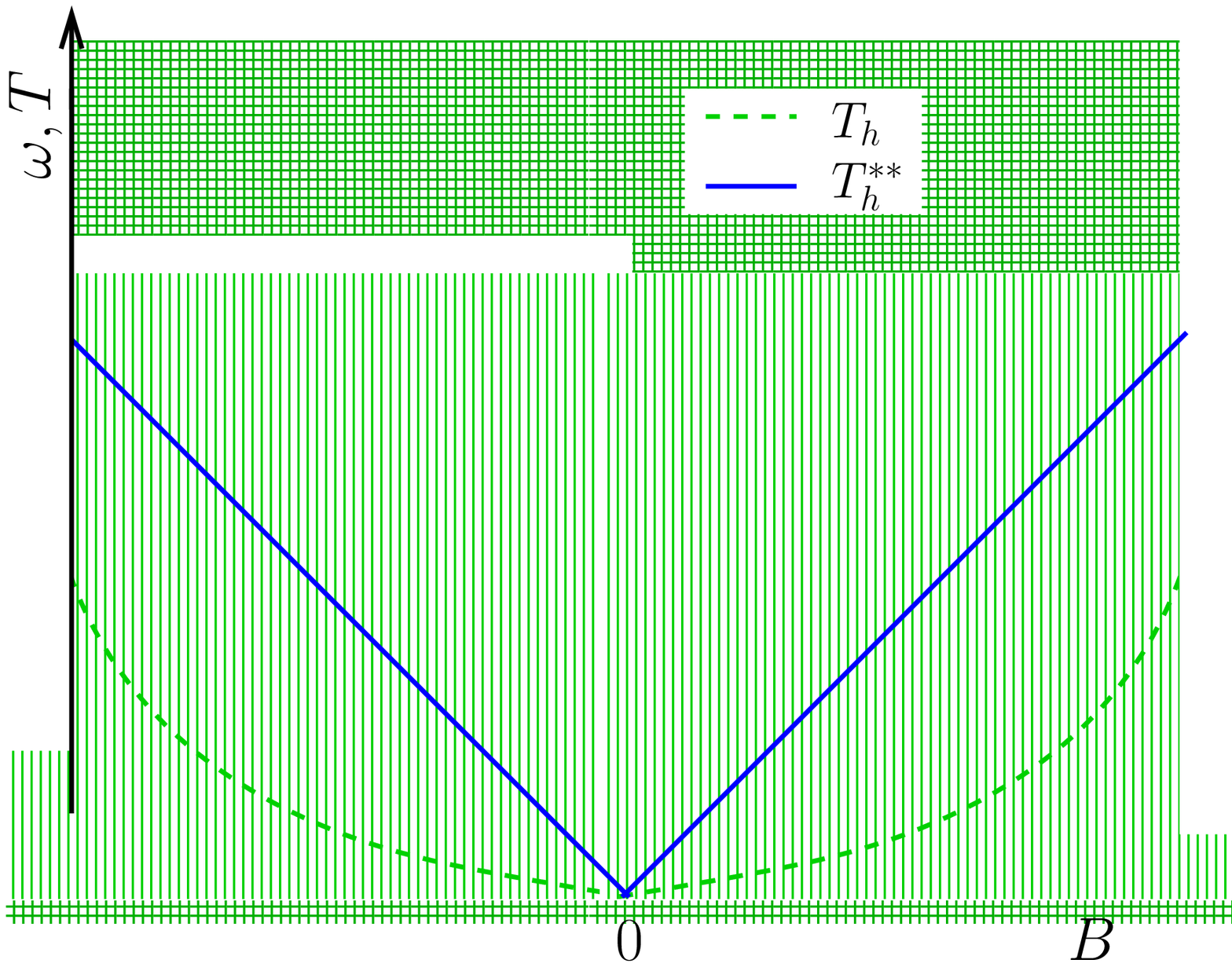}
  \caption{(color online) 
    Top: Sketch of the various 2CK scaling regimes in the presence of channel
    anisotropy for the local fermions bounded by $T^{**}_f$ from below and for the spin
    bounded by $T^{**}_s$, the crossover scale $T^{**}$ is also indicated.
    Bottom: Sketch of the 2CK scaling regime for the susceptibilities
    of the highest-weight fields bounded by $T^{**}_h$, and the crossover
    scale $T_h$ besides.
  }
  \label{fig:scaling_regimes}
\end{figure}

Our numerical calculations confirmed all our analytical expectations, and they
confirmed that actually, in the presence of an applied magnetic field, or
channel anisotropy, the  two-channel Kondo scaling regime is 
rather restricted, and it may also depend on the physical quantity considered.
In Fig.\ref{fig:scaling_regimes} we sketched the regimes where the 
pure two-channel Kondo behavior can be observed.  Notice that
in the presence of anisotropy the two-channel Kondo scaling regime of the 
 spin 
susceptibility has a boundary that differs from the 
boundary of the two-channel Kondo scaling regime of the $T$-matrix.

Some of the spectral functions show rather remarkable features: 
In a magnetic field, e.g., the  spectral function of the composite fermion, 
$F^\dagger_{\alpha,\downarrow}$  shows  a universal peak at a frequency 
$\omega = T_h$. This peak corresponds to spin-flip excitations 
of the impurity spin at the renormalized magnetic field. Remarkably, this 
peak is accompanied by a dip of the same size at the same frequency for 
spin down electrons. This dip is actually very surprising and is much harder to 
explain. Similar features appear but with opposite sign in the local fermions'
spectral functions. Even more surprisingly, this resonant 
feature is completely {\em absent} in the spectral function of the spin 
operators, $S^\pm$. 

One of the interesting results of our numerical analysis was that only the
composite superconductor ${\cal O}_{SC}$ has a logarithmically divergent
susceptibility. This is thus the primary candidate for  superconducting
ordering for a 2CK lattice system. We remark here that while for a single impurity the 
superconducting susceptibility seems to have a rather small amplitude, 
$\re\chi_{SC}\sim T_K/D_F^2 \ln(T_K/T)$, in a lattice model the mass of the carriers
is also renormalized, and therefore the bandwidth is expected to get  renormalized
as $D_F\to T_K$.\cite{Cox_97} As a result, the corresponding susceptibility
can be rather large, and drive, in principle, a superconducting instability. 
Interestingly, although the results are still somewhat controversial,\cite{Jarrell_96} 
in the two-channel Kondo lattice these local superconducting correlations  do
not seem to induce a superconducting transition.\cite{Anders_96} This may be, however, an artifact 
of the standard two-channel Kondo lattice model, which does not account 
properly for the orbital and band structure of an $f$-electron material.\cite{Coleman_98}
We believe, that in a more realistic lattice of 
two-channel Kondo impurities a composite superconducting order 
develops, similar to the one suggested in Ref.~\onlinecite{Coleman_98}. 
However, DMFT + DM-NRG calculations would be needed to confirm this belief.

Acknowledgement: We are especially grateful to L.\ Borda for making his code available 
for the Hilbert transformations and for the lot of valuable discussions. 
Useful comments on the manuscript from Z.\ Bajnok and I.\
Cseppk\"ovi are highly appreciated.
This research has been supported by Hungarian grants OTKA Nos. NF061726,
T046267, T046303, D048665, 
NK63066.

\appendix
\section{Scaling properties of two-point functions}
\label{app:scaling_prop}

In this appendix, we discuss the scaling properties of various scaling
functions. Essentially, we use the generalized  Callan-Symanzik  equations. 
For the sake of simplicity, let us first focus on the retarded Green's function of
the $z$-component of the operator $\phi_s$, 
\bea
{\cal G}\left(t,{\cal H}\right)&\equiv&
-i \langle\; [\phi_s^z(t),\phi_s^z(0)]\;\rangle_{{\cal
H}}\;\theta(t)\;,
\eea
and its Fourier transform, ${\cal G}(\omega,T)$. Let us investigate the
scaling properties of this function in the absence of magnetic field. 
From the fact that  $\phi_s$ is the field conjugate to the external ``magnetic field'', $h$, 
and that the partition function (generating function) must be scale invariant
under the renormalization group, 
 we easily get the following differential equation 
\bea\label{eq:greenf_scaling}
D \frac{\partial{\cal G}}{\partial D} + \sum_\mu \beta_{\mu}\frac{\partial{\cal
    G}}{\partial u_\mu}u_\mu + (2\beta_{h}-1){\cal G}\approx 0\;,
\eea
with $u_\mu$ a shorthand notation for the dimensionless couplings, 
$\{u_\mu\}=\{\kappa,\lambda,\dots\}$ that occur in  ${\cal
  H}$, 
 and $\beta_\mu$ the corresponding $\beta$-functions, 
\be 
\frac{{\rm d}\; {\rm ln }\;u_\mu}{{\rm d}x} = \beta_\mu\left( \{u_\nu\}\right)\;,
\ee
with $x=-{\rm ln}(D)$ the scaling variable.
In the vicinity of the two-channel Kondo fixed point the $\beta$-functions
just assume their fixed point value, which are just the renormalization
group eigenvalues, 
$y_\mu = d-x_\mu$, with the dimension $d=1$, since all operators are local and
live in time only.
 Since, for $ \phi_s$ we have $y_h= 1/2$, in the
close vicinity of the two-channel Kondo fixed point we obtain  
\be 
\frac {{\rm d} {\cal G}}{{\rm d}D  }\approx 0\;.
\ee 
One can also easily show that 
\be 
D  \frac {{\rm d} {\cal G}}{{\rm d} D   }
= - \omega \frac {{\rm d} {\cal G}}{{\rm d} \omega   }
\;.
\ee 
These relations  imply that,  ${\cal G}(\omega,T,D)$ is scale invariant, and
is only a function of $\omega/D$ and $T/D$. 
Clearly, similar equations hold for the correlation functions of all operators with
dimension $1/2$. Furthermore, the above scaling property can easily be modified for operators 
having dimensions $y_\mu\ne 1/2$.

\end{document}